Development and Validation of a Deep-Learning Model for Differential Treatment Benefit Prediction for Adults with Major Depressive Disorder Deployed in the Artificial Intelligence in Depression – Medication Enhancement (AID-ME) Study

Authors: David Benrimoh*[1,2], Caitrin Armstrong[2], Joseph Mehltretter[2], Robert Fratila[2], Kelly Perlman[1,2], Sonia Israel[2], Adam Kapelner[3], Sagar V. Parikh[4], Jordan F. Karp[5], Katherine Heller[6], Gustavo Turecki[1]

*Corresponding author

Affiliations:

1. McGill University, Department of Psychiatry
2. Aifred Health, Montreal, Canada
3. Department of Mathematics, Queens College, CUNY, Queens, NY
4. University of Michigan, Ann Arbor
5. Department of Psychiatry, College of Medicine-Tucson, University of Arizona. Tucson, AZ
6. Google, Mountain View, California

ABSTRACT:

**We introduce an artificial intelligence (AI) model aiming to personalize treatment in adult major depression, which was deployed in the Artificial Intelligence in Depression: Medication Enhancement (AID-ME) Study. Our objectives were to predict probabilities of remission across multiple pharmacological treatments, validate model predictions, and examine them for biases. Data from 9,042 adults with moderate to severe major depression from antidepressant clinical trials were standardized into a common framework and feature selection retained 25 clinical and demographic variables. Using Bayesian optimization, a deep learning model was trained on the training set and refined using the validation set. On the held-out test set, the model demonstrated an AUC of 0.65 and outperformed a null model (p = 0.01). The model demonstrated clinical utility, achieving an absolute improvement in population remission rate in hypothetical and actual improvement testing. While the model identified escitalopram as generally outperforming other drugs (consistent with the input data), there was otherwise significant variation in drug rankings. The model did not amplify potentially harmful biases. We demonstrate the first model capable of predicting outcomes for 10 treatments, intended to be used at or near the start of treatment to personalize treatment; AID-ME cluster randomized trial results are reported separately.**

## INTRODUCTION:

With over 300 million people affected worldwide [1], major depressive disorder (MDD) is a leading cause of disability. MDD poses an estimated socioeconomic burden of $326.2 billion per year in the U.S [2]. While several effective MDD treatments are available, only roughly one thirdof patients will remit with the first treatment they try[3]. This means that many patients must undergo a “trial and error” approach to treatment selection, which prolongs the duration of illness and leads to worse outcomes[4]. In order to improve outcomes in a cost- and time-effective manner, it would be useful to have a tool available which can help personalize treatment choice at the point of care, without requiring external imaging or genetic testing which can add time, expense, and complexity[5–7].

One can cast this challenge as one of pattern recognition: determining which patterns of patient features are predictive of outcomes with specific treatments. Machine learning (ML) methods are well-suited for this kind of pattern detection, and we have shown in previous work that deep learning ML models can be used to provide treatment-specific remission predictions and that they may provide advantages over other machine learning model types[5–8]. A recent review[9]of deep learning models for depression remission prediction found that some of the studies reviewed incorporated external testing, such as functional and structural magnetic resonance imaging, genetics, and epigenetics as features for the models. However, it remains infeasible to collect these measures routinely in clinical practice; we argue that a model that is to have a scalable impact should focus on measures that can be captured quickly and without specialized equipment, such as clinical and demographic variables[10]. One important limitation of most studies predicting remission in MDD is that they aim to differentiate outcomes between two treatments or predict the outcome of one treatment at a time[9,11]. This is not clinical reality; the reality is patients and clinicians must choose from over 20 antidepressants, psychological therapies, and neuromodulation treatments [12,13]. Another potential limitation of treatment selection machine learning models concerns the propagation of harmful biases (here defined as predictions which disadvantage a patient because of their socio-demographic background) [7,8,14]. In addition, clinicians may be apprehensive about the interpretability of the predictions of so-called “black box” machine learning models, obscuring the drivers of any given model-based recommendation[15]. Another important consideration is the data used to train models. Clinical trial datasets can prove difficult to work with because of the multitude of different symptom scales and outcome measures used across trials, as well as sample size and diversity limitations. On the other hand, data contained within electronic medical records requires that modelers contend with the biases inherent in observational data[16,17] as well as, in many cases, the lack of clear outcome measures.

These limitations have likely hindered the development and deployment of machine learning models into real clinical environments[8,17]. In previous work, we demonstrated deep learning models which can perform differential treatment benefit prediction for more than two treatments, which outperform other modelling approaches, and which are predicted to improve population remission rates[7,18,19]. We have also demonstrated methods for assessing model bias and for unifying clinical trial datasets in order to enable model training [8,19]. Finally, we have implemented versions of these models in simulation center and in-clinic feasibility studies in order to determine if clinicians and patients are accepting of these models, if they are easy to use in clinical practice, and if they are perceived by clinicians and patients to offer a clinical benefit [14,15,20,21]. Based on these initial studies, we developed an updated version of a differential treatment benefit prediction model which includes 10 pharmacological treatment options (8 first line antidepressants and 2 common combinations of first line antidepressants). This model, intended to be used by clinicians treating adult patients with MDD, was developed and validated using the methods we will describe below, and was then deployed as a static model in the Aifred Clinical Decision Support System (Aifred CDSS)[14,15,17,20,21] which served as the active intervention in the Artificial Intelligence in Depression – Medication Enhancement (AID-ME) study (NCT04655924), reported separately. Here we will report on the development and validation of the model used in this study (hereinafter referred to as the AID-ME model), in adherence to the (Transparent Reporting of a multivariable prediction model for Individual Prognosis Or Diagnosis) TRIPOD reporting guidelines for prediction model development and validation[22].

Our objectives herein are to 1) develop a model capable of predicting probabilities of remission across multiple pharmacological treatments for adults with at least moderate MDD and 2) validate model predictions and examine them for amplification of harmful biases. It is important to note that the results presented here could not be published prior to the conclusion of the AID-ME study in order to avoid potential unblinding of participants.

## METHODS:

We will now discuss the methods used to develop and validate the AID-ME model (see also [6–8]). Ethical approval was provided by the research ethics board of the Douglas Research Center.

### *Source of Data and Study Selection*

The data used to develop the AID-ME model was derived from clinical trials of antidepressant medications. Clinical trial data was selected primarily to reduce confounds due to uncontrolled clinician behavior which can change over time[23], biases in treatment selection or access to care[24,25]; to reduce data missingness; and, most importantly, because clinical trials contain clear clinical outcomes. All of the studies we used either randomized patients to treatment or assigned all patients to the same treatment (e.g. first level of STAR*D[3]), eliminating treatment selection bias operating at the individual clinician level. In addition, clinical trials have inclusion and exclusion criteria, which allow some assessment of potential selection biases which may be more covert in naturalistic datasets[26,27], though clinical trials are not free of selection bias[28]. Studies came from three primary sources: the National Institutes of Mental Health (NIMH) Data Archive; data provided from clinical researchers in academic settings; and data provided by GlaxoSmithKline and Eli Lilly by way of de-identified data requests administered through the clinical study data request (CSDR) platform. Results from a model trained on the pharmaceutical dataset are available[8].

Studies were selected in order to be representative of the intended use population: an adult population experiencing an acute episode of major depressive disorder, with demographics reflective of North America and Western Europe. These geographic regions were selected based on the availability of data and the sites likely to participate in the AID-ME study. These would be patients, much like those in the STAR*D dataset, who 1) could be seen in either primary or psychiatric care, 2) could have new onset or recurrent depression, and 3) would likely have other psychiatric comorbidities in addition to depression. However, the population would not be hospitalized, and would have primary MDD, and not depressive symptoms secondary to another medical condition, such as a stroke.

In order to achieve a set of studies that was consistent with this intended use population, we examined study protocols, publications, and clinical study reports. We excluded : 1) populations under age 18; 2) patients with bipolar depression/bipolar disorder; 3) studies in which the treatment of a major depressive episode was not the main objective (for example, studies of patients with dysthymia or with depressive symptoms without meeting criteria for an MDE in the context of fibromyalgia); 4) studies where the MDE is caused by another medical condition; and 5) studies of patients with only mild depression (though studies including patients with both mild and more severe depression were included, with mildly depressed patients later excluded). Given that the AID-ME study was planned to follow patients for 12 weeks, and that guidelines such as CANMAT suggest assessing remission after 6-8 weeks[13], we included studies which had lengths between 6 and 14 weeks.

Studies were conducted between 1991 and 2016. After working to secure as many studies as possible, we began with 57 studies for consideration. After reducing the number of studies based on the criteria noted above, as well as excluding some studies to ensure that specific drugs were not excessively over-represented, we were left with 22 studies. See PRISMA diagram (Fig 1). Important guidance in the selection of treatments of depression is provided by the Canadian Network for Mood and Anxiety (CANMAT) treatment guidelines for MDD in adults[13]. These CANMAT guidelines were used to identify treatments to be included in the model and were referred to during interpretation of model results. Only patients with in arms with medication doses reaching minimum effective doses as per the CANMAT guidelines were included.

*Participants*

Participants were all aged 18 or older and were of both sexes. They were treated in primary care or psychiatric settings. Participants were excluded if, at study baseline they had a depression of less than moderate severity based on the study's depression rating tool. They were also excluded if they were in their respective study for less than two weeks, given that it is unlikely that an adequate determination of remission or non-remission could be made in this timeframe. Patients who remained in their studies for at least two weeks were included even if they dropped out of the study early.

*Treatments*

While a number of treatments were used across the studies included, we could only include treatments in the final model if there were sufficient observations of the treatment available for the model to learn to predict its outcome. In line with our previous work, we did not generate predictions for treatments provided to fewer than 100 patients in the dataset[7]. In addition, participants taking less than the dose range recommended in the CANMAT guidelines [13] were also eliminated from the dataset, as the CANMAT guidelines formed an integral part of the AID-ME study. As our modeling technique does not depend on a placebo comparator and no placebo treatment was to be recommended by the model in clinical practice, patients who received placebo were not included. The included medications are noted in Table 1. These include a diverse array of commonly used first-line treatments as well as two combinations of first line medications which are commonly used as augmentation strategies [29]. After these exclusions, 9042 patients were included in the final analysis across both datasets. Patients were not excluded if they received other non-pharmacological interventions; the variability of available information in our data made it impossible to fully account for non-pharmacological treatments, but this variability is also in line with what is seen in clinical practice.

*Outcome*

The predicted outcome was remission, cast as a binary variable so that the model could predict remission or non-remission as a label, and provide a probability of remission. Remission was chosen as it is the gold-standard outcome in treatment guidelines; achieving remission is important because the presence of significant residual symptoms increases relapse risk[13]. Remission was derived from treatment scores on standardized questionnaires and binarized based on cutoffs derived from the literature. These included the Montgomery-Asberg Depression rating scale (MADRS; remission defined as a study exit score < 11)[30–32], the Quick Inventory of Depression Scale - Self Report (QIDS-SR-16; remission defined as a score < 6) and the Hamilton Depression Rating Scale (HAMD; remission defined as a score of < 8) [30,33]. Remission was measured at the latest available measurement for each patient in order to preserve the most information for each patient (such that patients who remit but then become more symptomatic are not incorrectly classified as being in remission)[7]. Given this approach, and the fact that all patients were in their respective studies for at least two weeks, no patients were missing outcomes and outcomes were not imputed. Model developers and assessors were not blinded to the selected outcome given the limited size of the team and the need to carefully specify remission variables and cutoffs; however, final testing on the test set was carried out by a team member who was not involved in the development of the AID-ME model.

*Data preprocessing and creation of stratified data splits*

A full method description for the transformation pipeline and quality control measures is available in [8]. Briefly, we created a “transformation pipeline” to combine different questions across datasets to prepare as input variables for feature selection. First, we developed a custom taxonomy inspired by the HiTOP taxonomy system [34] to categorize the study questionnaire data across clinical and sociodemographic dimensions. Standard versions of each question were tagged with at least one taxonomic category which allowed for items with the same semantic meaning to be grouped together and created into a “transformed question” representative of this category. Questions with categorical response values were grouped together and values were scaled with either linear equating or equipercentile equating, depending on the response value distributions [35]. No transformation was required if all response values were binary, but if there was a mix of categorical and binary questions, the categorical values were “binarized” to either 0 or 1 based on the response value text and how it compared to the response value text of the natively binary questions.

To address potential inconsistencies in input data ranges,which is crucial to consider for neural networks(Sola and Sevilla 1997) we applied a standardization process during data preprocessing. After equipercentile equating transformed responses into a unified standard across similar question groups, we standardized all input features to ensure consistency in scale and range. This approach mitigates the risk of variability in feature magnitudes negatively impacting model performance. Additionally, for features with imputed values, we added a small constant uniformly to avoid zero-valued inputs, which can destabilize the training process. This step further supported stable and effective model training. The standardization ensures that all input features are appropriately scaled for the model, aligning with best practices for neural networks(Novak et al. 2018).

To generate our data splits, we stratified each division using the binary remission and treatment variables. The sequential splitting method was used to generate the three data splits: utilizing the train_test_split function from scikit-learn [36], we first separated our aggregate dataset into a test set with a 90-10 stratified split. The leftover 90% was further divided to form the train and validation sets, comprising 80% and 10%, respectively, of the original data. This process yielded an 80-10-10 split for our datasets. The test set was held out and not used in any way during model training; it was only used for testing after the final model was selected.

*Missing data*

We handled missing data as follows. First, features with excessive missingness (over 50%; see [8]) were removed from the dataset. After this data was imputed using multiple imputation by chained equations (MICE) provided by the package Autoimpute[37], which was built based on the R *mice* function. To reduce the number of variables available to feed into each variable's imputation, we used multicollinearity analysis [38,39]. The variance inflation factor of each variable was used to measure the ratio of the variance of the entire model to the variance of a model with only that specific feature. The threshold for exclusion was set at 5, the default. This enabled us to have a reduced set of variables to feed into the MICE algorithm without extreme multicollinearity as would be expected given the number of available variables. To use the MICE algorithm we generated 5 datasets and took the average between all datasets for the variables using the least-squares or the predictive mean matching (PMM) strategy (continuous variables) and the mode for binary or ordinal variables[40]. Race/ethnicity, sex, treatment, and outcome (remission) were not imputed; all patients included had data for these variables.

In order to facilitate the deployment and interpretation of our model, we chose to pursue a single-dataset strategy when it came to imputation. However, it is commonly recognized that capturing the variability inherent in imputing a single variable during multiple imputation is important. Our approach therefore aimed to compromise between a standard multiple imputation approach and a single imputation approach by averaging our imputed datasets, as was done in a recent paper examining dementia data(McCombe et al. 2022).

*Feature selection*

As a neural network was used for the classification model, we chose to also use neural networks for feature selection [41]. To accomplish this we implemented a new layer called CancelOut [42]. CancelOut is a fully-connected layer, allowing us to create a classification model with the same task of training with a target of remission. The CancelOut layer has a custom loss function that works as a scoring method so that by the end of training we can view and select features based on their score. The retained features were input into our Bbayesian optimization framework, which was used for selection of the optimal hyperparameters.

*Final Predictors*

Predictors, also known as features, were clinical and demographic variables which were selected by the feature selection process. These features are listed in Table 1. The team developing and validating the model was not blinded to features used for outcome prediction.

Nominally, 1415 predictors were available in the data provided by the NIMH and other researchers, and 624 predictors were available in the data derived from the pharmaceutical databases. However, only 183 of these features were available in common across these data sources. Furthermore, after removing variables with excess missingness, 41 features remained. After running these predictors through our feature selection process during model development, 26 features remained (Table 1).

*Model Development*

A deep learning model was prepared by arranging a set of hyperparameters (including, for example, the number of layers, number of nodes per layer, activation function, learning rate, number of training epochs, etc). During the training process, the model's hyperparameters were refined using a Bayesian optimization process (BoTorch package)[43]. The number of layers and the Bayesian optimization targets (AUC and F1 score in the validation data) were provided to the Bayesian optimization in a fixed manner. Early stopping, within standard ranges for epoch number between 150 and 300, was used to prevent overfitting[44]. Further details are available under Code Availability and in the supplementary material. Our modelling pipeline additionally relied on a series of open source Python packages which are detailed in the supplementary material.

Models were trained and their performance was optimized using the validation dataset. The best performing architectures were selected to produce a final model. This final model was then run on the test set. This test set was not used in any way during model training. See supplementary material for further details.

Selection of deep learning as the modelling approach

While deep learning has been foundational to the success of many recent advances in computer science (such as computer vision), its utility in comparatively simple datasets, such as the tabular dataset used here, compared to more traditional approaches is not guaranteed (see (Hollmann et al. 2025) for discussion). In this paper our objective is to report the results of the version of our model which was included in the AID-ME trial. However, in previous work, we have demonstrated that deep learning approaches can provide advantages with the data used here. In a comparison with a gradient boosting approach that used elastic net for feature selection(Chekroud et al. 2016), we have shown that a deep learning approach can provide roughly equivalent performance while improving generalizability of predictions(Mehltretter et al. 2020; Mehltretter et al. 2019). In more recent work using the same approach adopted in this paper, we demonstrated that our deep learning approach improves upon results from a logistic regression model(Perlman et al. 2024). This, in addition to the potential for deep learning to incorporate more complex data types(Benrimoh et al. 2018) (such as images, other biosignals, or natural language) in future work, led us to pursue a deep learning approach for the current model.

*Model Usage After Freezing*

When the model is used for a new patient after training and validation is completed and the model is made static, data provided for a patient for the selected features is inputted into a forward pass of the model, and run for each of the possible medications, generating a remission probability for each treatment. We additionally use the saliency map algorithm to generate the interpretability report that provides the top 5 most salient variables for each inference performed[7]. Finally, we calculate the average remission probabilities across all predicted treatments. The results are then packaged and returned to the physician to make their determination (see below and Figure 5). During clinical use, data is provided by patient via self-report questionnaires and by clinicians using a clinician questionnaire with built-in instructions, meaning that users require only minimal training to provide the data. Missing responses are not permitted and so no missing data is provided to the model when a new patient uses it.

*Primary metrics*

The model evaluation was primarily driven by optimizations across the Area under the Receiver Operating Curve (AUC), as well as the F1 score (where remission was predicted if the model estimated a remission probability of 50% or greater). As the AUC is scale- and classification-threshold-invariant, it provides a holistic and well-rounded view of model performance (Huang, 2005). Since the F1 Score is the harmonic mean of precision and recall it helps us assess our model's handling of imbalanced data (Raschka, 2014). In addition, Accuracy, Positive Predictive Value (PPV), Negative Predictive Value (NPV), Sensitivity, and Specificity were also assessed, as these are commonly utilized in clinical studies and are meaningful to both machine learning engineers and clinicians. NPV and PPV especially are important in helping clinicians understand the value of positive and negative predictions.

Another important metric was model calibration (specifically expected and maximum calibration error)[45]. This metric reflects the relationship between the outcome predictions of the candidate model and the observed outcomes of the grouped sample. A model is said to be well calibrated if, for a group of patients with a mean predicted remission rate of x%, close to x% individuals have reached remission.

There was no difference in metrics used to assess the model during training and testing. A pre-specified model performance of an AUC of at least 0.65 was set as a target. Further discussion of pre-specified metrics is available in the supplementary material.

*Secondary Metrics*

To estimate clinical utility we use two analysis techniques: the “naive” and “conservative” analyses, which are described in detail in [7] and in the supplementary section. The naive analysis estimates the improvement in population remission rates, by taking the mean of the highest predicted remission rate for each patient from the test set from among the 10 drugs, and taking the mean of these predicted remission rates. The conservative analysis only examines predicted and actual remission rates for patients who actually received optimal treatment, utilizing a bootstrap procedure performed on the combined training and validation sets[46]. Both metrics allow estimation of predicted improvements in population remission rate, and we set a 5% minimum absolute increase in population remission rate, consistent with the benefit provided by other decision support systems such as pharmacogenetics [47].

*Interpretability Report Generation*

We generated interpretability reports using the saliency method [48] to assess the importance of the input data for each prediction produced. Using GuidedBackprop, we generated a numerical output that designates the importance of a feature in determining the output[49], providing a list of prediction-specific important features which are similar to lists of ‘risk’ and ‘protective’ factors which clinicians are used to. The top 5 features in this list are provided to clinicians[7] (see Fig 6 for an example).

*Bias Testing*

We utilized a bias testing procedure we have previously described in [8] to determine if the model has learned to amplify harmful biases. For race, sex and age groupings, we create a bar graph depicting true remission rates for each group, as well as the mean predicted remission probability for that group. This allows us to assess if the model has learned to propagate any harmful biases, and to examine disparities in outcomes based on sociodemographics in the raw data (e.g. worse outcomes for non-caucasian groups). We also verify the model’s performance with respect to the observed remission rates for each treatment. The subgroups for age were 18–25.9;26–40.9; 41–64.9; 65+. The subgroups for race/ethnicity were Caucasian, African Descent, Asian, Hispanic, Other. Finally, subgroups for sex were male and female.

*Sensitivity Analyses*

Sensitivity analysis was used to probe model behavior in response to manipulation of specific variables to determine if the model responds in a manner that is consistent with previous evidence. For example, we can plot how the remission probability changes in response to artificially increasing suicidality; the model should produce a clear trend towards lower remission probabilities as suicidality increases[10].

RESULTS:

*Participants*

The final analysis dataset included 9,042 participants. Their demographics, by data split (train, test, validation), are presented in Table 2. As expected with depression statistics, the sample was predominantly female (63.6%)[50]. The remission rate of the entire sample was 43.2%, reasonably in line with what would be expected in clinical trial data for remission in depression[51]. The data splits were stratified in order to ensure that remission rate and percentage allocated to each drug was similar between the splits. As noted, overall data missingness is described in the supplementary material. Due to the data preprocessing employed, no participants included in this dataset were missing outcomes.

*Final model characteristics*

Our Bayesian optimization procedure identified an optimal model featuring two fully-connected hidden layers, each with 40 nodes, leveraging exponential linear units (ELU) and a dropout rate of 0.15 to enhance generalization. The model's prediction layer utilized the softmax function to calculate remission probabilities. The Adam optimization algorithm, set at a learning rate of 0.001, was employed for adjusting network parameters during the training phase. To avoid overfitting, the model incorporated an early stopping mechanism. The maximum training duration was set at 300 epochs, with a patience threshold of 100 for early stopping. The model completed the full 300 epochs of training.

*Model performance*

The model performance metrics are provided in Table 3 for the training, validation and test sets. As can be seen, the model generalized well to the test set (AUC is similar in the test and training set). Overall accuracy and AUC are somewhat low, but in line with results of other models in these kinds of datasets (see discussion below). In Figure 2, we present the calibration plot for the test set; plots for the other sets are available in the supplementary material. The model was reasonably well calibrated, with an Expected Calibration Error (ECE) of 0.034 and a Maximum Calibration Error (MCE) of 0.087 (lower values are better). The model outperformed a null model (which always predicts non-remission, the modal outcome) on the test set ($p$ = 0.01).

In Supplementary Tables 5 and 6 we present the frequency with which each drug was ranked in each possible place (from 1st to 10th). As these tables demonstrate, the model has learned to most often predict the highest remission rate for escitalopram. This is consistent with the input data (See Figure 3) wherein escitalopram had the highest remission rate of all drugs. This is also consistent with the CANMAT guidelines, which note escitalopram among the more effective first-line antidepressants, and with a previous metanalysis which had similar treatments included [52] and found that escitalopram had among the highest effectiveness ranking and superior tolerability compared to other antidepressants (we did not have access to data regarding mirtazapine monotherapy). However, as S5 and S6 demonstrate, beyond escitalopram there was significant variation in drug rankings, and at the aggregate level the rankings are reasonably consistent with previous metanalyses[12,52]. In addition, we provide in the supplementary material a listing of model performance for each drug.

As for the assessment of clinical utility, the naïve analysis suggest an absolute improvement in population remission rate from 43.15% to 53.99% on the test data. The average difference between the remission rate predicted for the best and worst drug, per patient, is 10.8, suggesting a widespread between treatments. There was additionally an absolute improvement from 43.21 to 55.08% improvement under the conservative analysis on the testing data (when considering real remission rates). See Table 4 for numerical results of both analyses. The bootstrap sampling done in these analyses allows us to determine if model-directed treatment was significantly better than random allocation; improvement in predicted remission rates for those receiving model-optimal treatment was statistically significant ($p < 0.001$) using the procedure from[46]. See supplementary material for further details.

*Model bias assessment and sensitivity analysis*

*In Figure 3, we present results for our analysis aimed at identifying if the model has learned to amplify harmful biases. Figure 3 demonstrates, for subgroups relating to age, sex, and race/ethnicity, the observed and predicted remission rates for each subgroup. As the figure shows, for all subgroups for which more than a small number of patients were available, there are no cases in which the model predicts a significantly worse outcome (>5% below observed rate8) than is observed. In previous work(Perlman et al. 2024), we argued that an under-prediction of remission rate would be potentially harmful for a subgroup, and this analysis was intended to identify such under-predictions.*

In Figure 4, we present example results from our sensitivity analyses. These show remission rates decreasing with increasing severity of baseline suicidal ideation and increasing with the severity of weight loss. These are consistent with previous literature in which increased suicidality has been linked to increased severity of illness and a reduction in treatment response [53,54]; in addition, decreased weight as a symptom has previously been linked to an increase in antidepressant treatment response with fluoxetine [55].

*Model interface*

Once we completed model testing and validation, the final model was saved in a frozen state to prevent further updating and uploaded into the Aifred CDSS platform. This platform will be fully described in the accompanying clinical trial results paper. In Figure 5 and Box 1, we demonstrate how the results of the model are provided to clinicians. Probabilities, rather than class labels, are provided in order to avoid being overly prescriptive and to encourage clinicians to think of the prediction as ‘one more piece of information’ rather than as a directive[56].

DISCUSSION:

In this paper we have presented results for a differential treatment prediction model of 10 antidepressant treatments, using solely clinical and demographic features. While combining clinical and demographic variables with other modalities, such as genetic information or imaging has previously demonstrated superior performance[9], the clinical reality, in the majority of settings in which depression is treated, is that clinicians and patients will likely only have access to simple clinical and demographic information. As such, one of the main strengths of our model is that it can be easily used in any setting where an internet connection is available, with results available as soon as the straightforward questionnaire is completed and without requiring equipment other than a computer or mobile phone. This model was generated in order to be used in a clinical trial which has recently concluded (NCT04655924).

As this is one of the first treatment outcome prediction models to be deployed in a clinical environment in psychiatry, we carefully selected the decision for which the model could be of assistance, in order to avoid a situation where an incorrect model prediction could cause significant harm. For this reason, we elected to help differentiate between commonly used first line treatments. When the model makes accurate predictions, it can help improve treatment success rates and assist clinicians and patients in making decisions between the many available treatments; when the model makes incorrect predictions, patients still receive first line treatments for depression which are known to be safe and effective. This effort to mitigate risk via a careful selection of use-case notwithstanding, we attempted to examine our model thoroughly to ensure that it had the most potential to provide benefit while avoiding possible harm.

In order to assess potential benefits, we assessed potential clinical utility by determining projected improvements in remission rates and by assessing the extent the model was able to differentiate between treatments. We were able to demonstrate model metrics and potential improvements in remission rates generally consistent with our and others' previous work[6,7,57]. We note some reduction in AUC compared to our previous models, likely related to the reduced feature set available as a result of merging together a large number of datasets, resulting in reduced feature availability (see below).The benefit of including this large array of datasets, however, was to allow us to make predictions for 10 different individual treatments, more than any model in MDD which we are aware of.  Given the large number of treatments available, sufficient treatment coverage is crucial in order to produce a clinically useful treatment prediction model. Importantly, the model did develop a preference for escitalopram as the most effective drug and this drug is thus a main driver of the predicted improvements in population remission rate.

While the preference for escitalopram (Tables S5, S6) is consistent with both the input data and previous meta-analysis,[12,52] one might be concerned about the potential impact on clinician decision-making. However it is important to note that the treatments aside from escitalopram show significant variation, such that clinicians and patients, should they decide escitalopram is an inappropriate option because of cost, medication interactions, or having previously tried the medication, would likely benefit from the differential prediction between the other 9 treatments. Another beneficial aspect of our model is its interpretability report, which has in previous work been shown to support clinician trust in model predictions and, in turn, use of these predictions in treatment decisions in a simulation study[15].

We recognize that, likely due to systemic factors(Lesser et al. 2007; Fortuna et al. 2010), certain subgroups may have lower observed rates of remission; this is true in the data available, and can be seen in the observed remission rates noted in Figure 3. In order to assess potential harms of our model, we examined the performance of the model in different sub-groups to ensure that it did not amplify these existing biases. This is especially important given, as we note below, that our dataset was missing important socioeconomic information which may help explain differences in subgroup outcomes(Lesser et al. 2007). As noted, we did not find any signs of amplified negative bias. While this holds true in the intended use population, it is important to note the model would need to be re-validated and potentially re-trained if used in different populations. In addition, we demonstrated model validity not only by assessing performance metrics, but also by determining, through sensitivity analyses, that associations learned by the model were consistent with previous literature.

With respect to the chosen interpretability method (example results of which are available in Figure 5), we did not want to obscure the complex nonlinear relationships that a deep learning model would harness with a method that linearizes the interactions (e.g., SHAP(Lundberg and Lee 2017)). To avoid the process of linear approximations, we chose to employ a gradient based method, GuidedBackProp(Springenberg et al. 2014), for per-prediction feature importance. To better understand global feature interactions, we employed sensitivity analysis to better understand the directionality and magnitude of prediction effects for each feature. In future experimentation, we plan to explore combinations of methods such as DeepLift(Shrikumar et al. 2017) or Generalized DeepSHAP(Chen et al. 2022).

There are a number of significant limitations to the present work. The majority of these limitations are related to the quantity and quality of data available. The limitations of our approach to merging data across studies are discussed here[8]. There was a significant amount of missing data, and as such predictions can only be as valid as the imputation procedure (see[8] for further discussion). The data used to train the model, while large in the context of psychiatric research, did have some important lacunae. For example, it did not include all first line treatments (e.g. vortioxetine, mirtazapine monotherapy). This data was not available to us, and if it was, it would likely have resulted in a model with more varied options for the most effective treatment as shown in meta-analyses [12,13,52].

In addition, the dataset did not contain a great number of socio-demographic variables, which we have previously shown to be predictive of remission [6,7]. Our model would likely have improved performance if more features were available. Additionally, we included data mostly from North America and Western Europe, due to data availability as well as our intent to use the models in these populations; future work would be needed to further validate the model in other populations. Finally, our data only covers initial treatment choice and is not capable of adjusting predictions after treatment failure - an important area to explore in future work. It is important to consider that one of the strengths of AI systems is in their ability to learn and be updated over time as more data becomes available; while we argue that the current model has clinical utility, clinicians involved in the clinical testing of our previous models have noted their interest in seeing utility improve over time in successive versions of the model.

While this paper focuses on the documentation of the development and validation process for the model used in the AID-ME trial, future extensions or updates to this model would benefit from thorough benchmarking against both traditional models, such as gradient boosting and regression models, as well as against novel model types which were not available at the time of our model's creation, such as transformer-based foundation models adapted to tabular data(Hollmann et al. 2025).

We have demonstrated that it is possible to create models which can provide differential treatment benefit predictions for a large number of treatments, without requiring complex or expensive biological testing. We present a process by which such models can be developed and validated, and describe the manner in which their results can be presented to clinicians. This model has since been field-tested in a clinical trial, representing, we hope, a step forward for the use of predictive tools in mental health clinical practice. Future focus on improving the breadth of features available for machine learning model training will likely further accelerate the clinical utility of machine learning models. Similar approaches could potentially be used to assist decision making in other illnesses.

FIGURES AND TABLES:

- Figure 1 PRISMA diagram for study inclusion and included medications

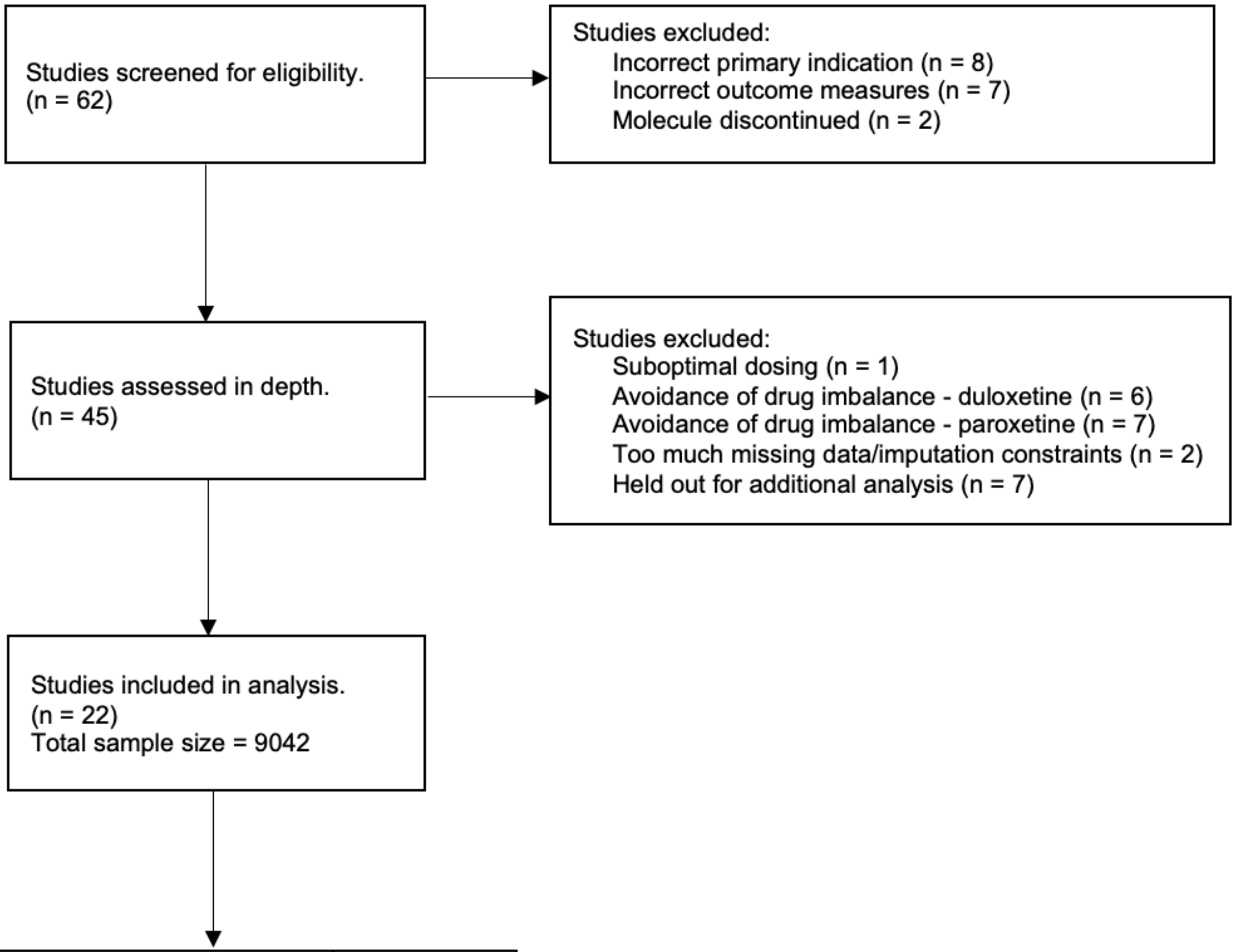

| **Treatment** | **n** |
|---|---|
| Citalopram | 2311 |
| Paroxetine | 1502 |
| Duloxetine | 1428 |
| Venlafaxine | 988 |
| Fluoxetine | 662 |
| Bupropion | 628 |
| Sertraline | 606 |
| Escitalopram | 544 |
| Venlafaxine-XR + Mirtazapine | 188 |
| Escitalopram+Bupropion | 185 |

Table 1: Final features. The data type is numeric unless otherwise specified.

| **Feature name** | **Type/Levels** |
|---|---|
| Age | Integer |
| Anxiety – Psychic | Likert |
| Anxiety – Somatic (categories: Gastrointestinal, indigestion, Cardiovascular, palpitation, Headaches, Respiratory, Genito-urinary, etc.) | Likert |
| Anxiety/Tension/Inability to relax – binary | Categorical (binary) |
| Anxious state | Likert |
| Decreased appetite | Likert |
| Early insomnia | Likert |
| Excessive guilt – binary | Categorical (binary) |
| Feelings of worthlessness | Likert |
| Gastrointestinal disturbances | Likert |
| Genital symptoms (Loss of libido, menstrual disturbances) | Likert |
| Guilt | Likert |
| Hopeless outlook on future | Likert |
| Hypochondriasis | Likert |
| Insight | Likert |
| Late insomnia | Likert |
| Leaden paralysis | Likert |
| Middle insomnia | Likert |
| Negative thoughts – binary | Likert |
| Overall suicidal ideation | Likert |
| Psychomotor agitation | Likert |
| Race/ethnicity | Categorical (African Descent, Asian, Caucasian, Hispanic, or Other (e.g., Indigenous / American Indian, Hawaiian or Pacific Islander, More than one race) |
| Sex - binary | Categorical (Male, Female) |
| Suicidal ideation and planning – binary | Categorical (binary) |
| Treatment Variable | Categorical |
| Weight loss | Likert |

Table 2: Demographics for each data split

| | African Descent | Caucasian | Hispanic | Other | Asian | Null Race | Male | Female | Age (18-25.999) | Age (26-40.999) | Age (41-64.999) | Age (65+) | Remission | Total n** |
|---|---|---|---|---|---|---|---|---|---|---|---|---|---|---|
| Train* | 575 (7.82%) | 5584 (75.97%) | 142 (1.93%) | 136 (1.85%) | 692 (9.41%) | 221 (3%) | 2672 (36.35%) | 4678 (63.64%) | 799 (10.87%) | 2391 (32.53%) | 3543 (48.2%) | 617 (8.39%) | 3,173 (43.17%) | 7,350 (81.28%) |
| Test* | 57 (6.5%) | 677 (77.37%) | 22 (2.52%) | 13 (1.48%) | 77 (8.8%) | 29 (3.3%) | 323 (36.91%) | 552 (63.1%) | 97 (11.09%) | 263 (30.1%) | 445 (50.86%) | 70 (8%) | 380 (43.43%) | 875 (9.67% |
| Validation* | 64 (7.83%) | 633 (77.47%) | 21 (2.57%) | 16 (1.96%) | 60 (7.34%) | 23 (2.82%) | 300 (36.72%) | 517 (63.28%) | 78 (9.55%) | 249 (30.48%) | 419 (51.29%) | 71 (8.69%) | 349 (42.71%) | 817 (9.03%) |
| Total** | 696 (7.7%) | 6,894 (76.24%) | 185 (2.05% | 165 (1.82%) | 829 (9.17) | 273 (3.02%) | 3,295 (36.44%) | 5,747 (63.56%) | 974 (10.77%) | 2,903 (32.11%) | 4,407 (48.74% | 758 (8.38%) | 3,902 (43.15%) | 9,042 (100%) |

*-Percentages calculated with a denominator of the sub-total n per dataset split

**-Percentages calculated with a denominator of the full dataset

Table 3: Final Model Accuracy Results. Columns 4-8 results are based on classifying remission if the predicted remission probability is 50% or greater.

| Data Source | Accuracy | AUC | F1-Score | Sensitivity | Specificity | NPV | PPV |
|---|---|---|---|---|---|---|---|
| Validation Set | 0.60 | 0.62 | 0.55 | 0.56 | 0.62 | 0.65 | 0.53 |
| Held-Out Test Set | 0.60 | 0.65 | 0.51 | 0.47 | 0.70 | 0.55 | 0.64 |

Table 4: Results of Naive and Conservative Testing

| Test | Sample N on which test was conducted | Baseline remission rate before model | Remission rate as per the test | Absolute difference | Relative difference (Absolute difference / baseline) | Average difference (%) between best and worst drug for each patient |
|---|---|---|---|---|---|---|
| Naive | 875 | 43.63 | 53.98 | 10.35 | 0.24 | 10.8 |
| Conservative-predicted remission | 8167[i] | 43.21* | 57.41* | 14.2* | 0.33* | n/a** |
| Conservative- actual remission | 8167[i] | 43.21* | 55.08* | 11.87* | 0.28* | n/a** |

*Rates are based on the average of the samples taken for our bootstrap (n = 100) analysis.

**Analysis not carried out given the large number of folds

[i]Note: bootstrap analysis was carried out on the combined train and validation sets.

Figure 2 Test Set Calibration Plots

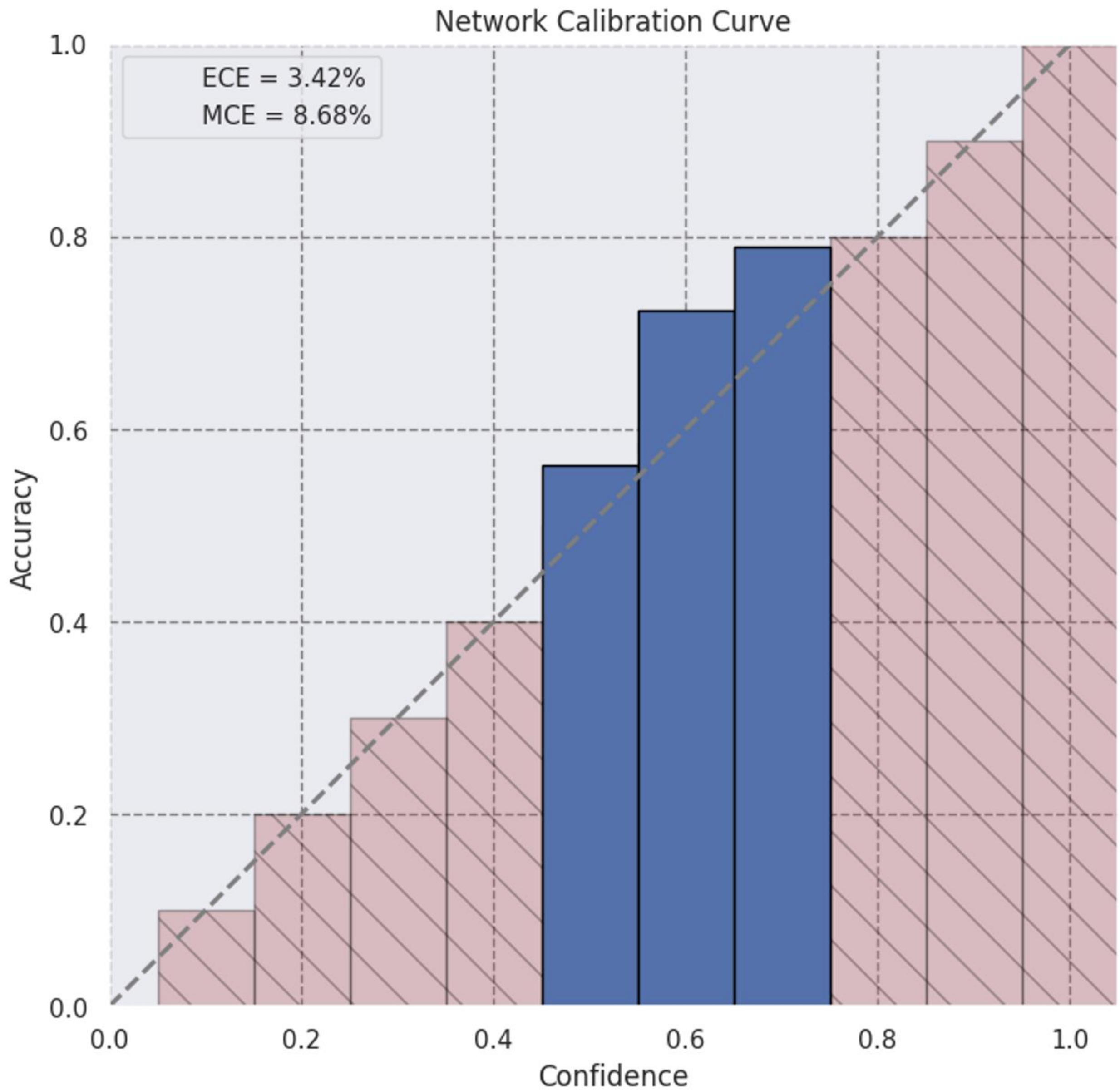

Calibration plot for the test set. ECE = expected calibration error. MCE = maximum calibration error.

Figure 3 Bias Plots

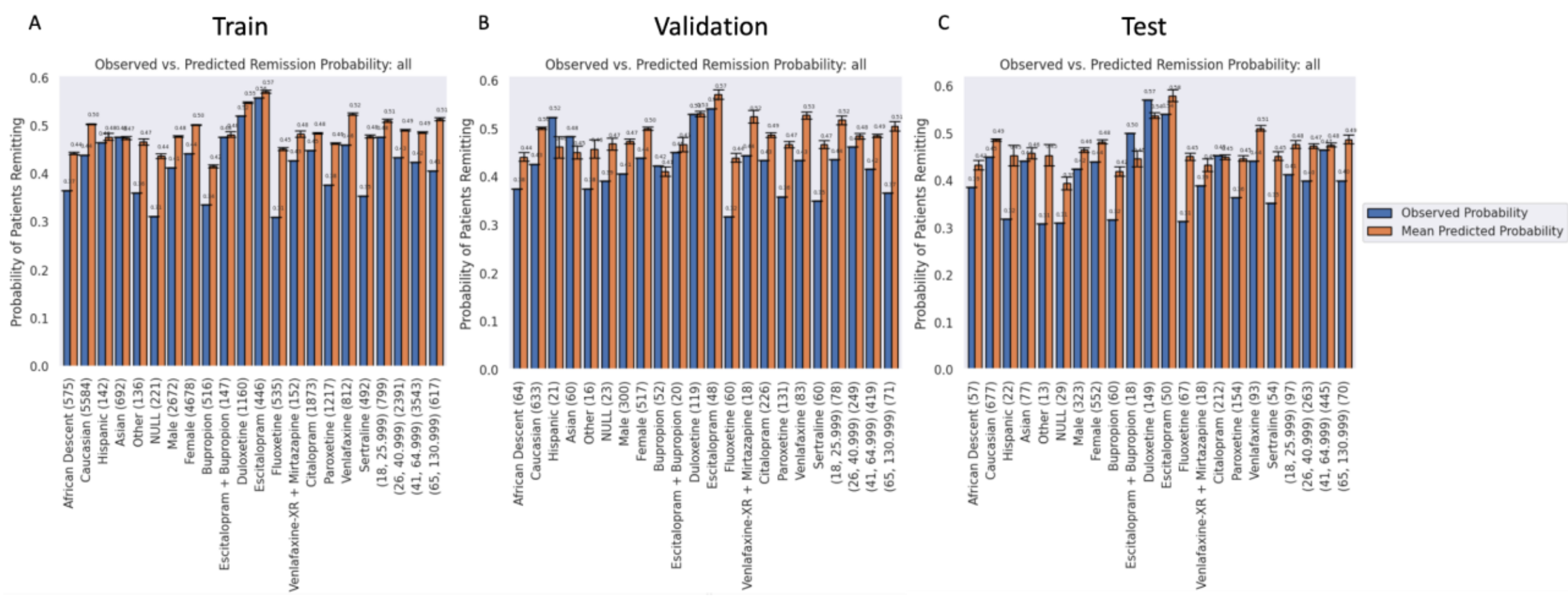

Plots of observed and predicted remission probability for drug, sex, ethnicity, and age groups (extreme right). Error bars represent standard error.

Fig 4: Selected Sensitivity Plots- Test Set

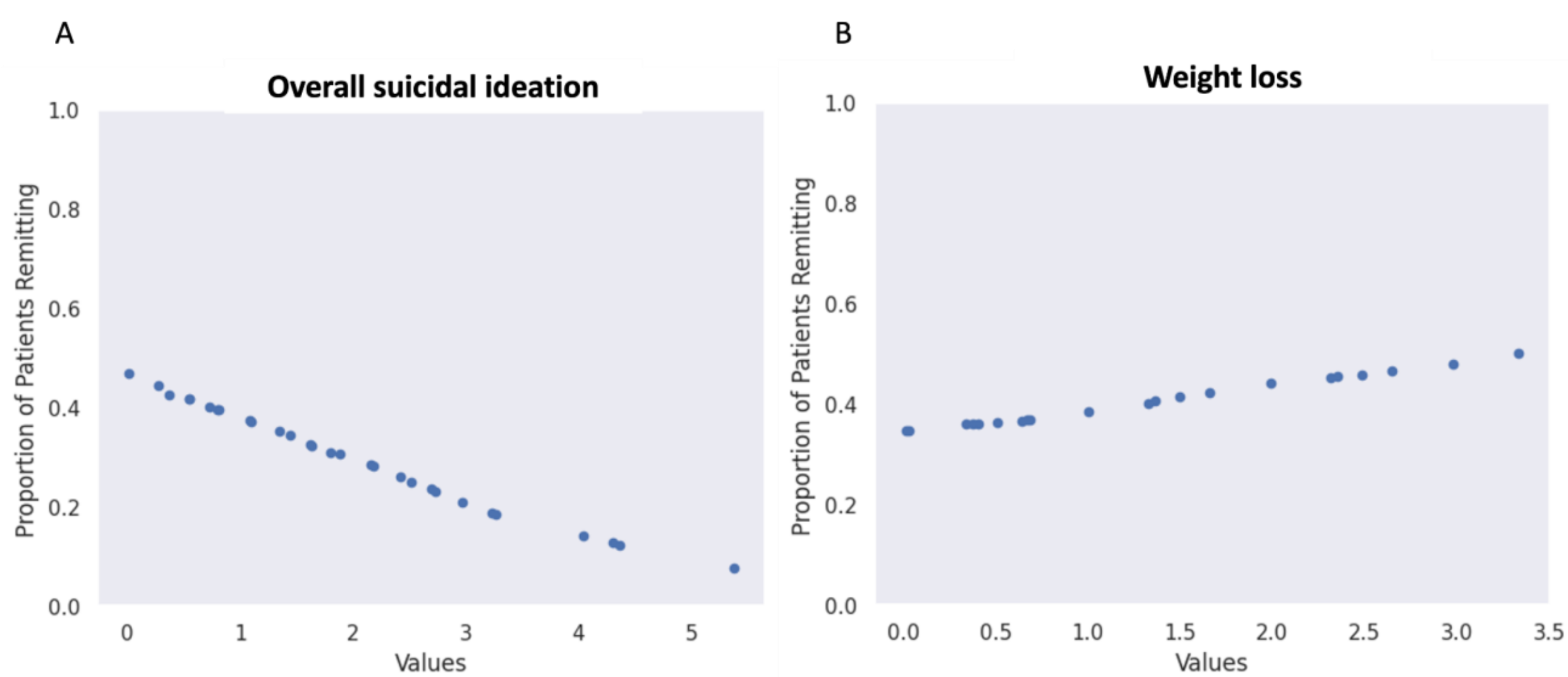

Plots of remission rate predicted by the model as each variable (suicidal ideation in A and baseline weight loss in B) at baseline is increased.

Fig 5: example output from the Clinical Decision Support System

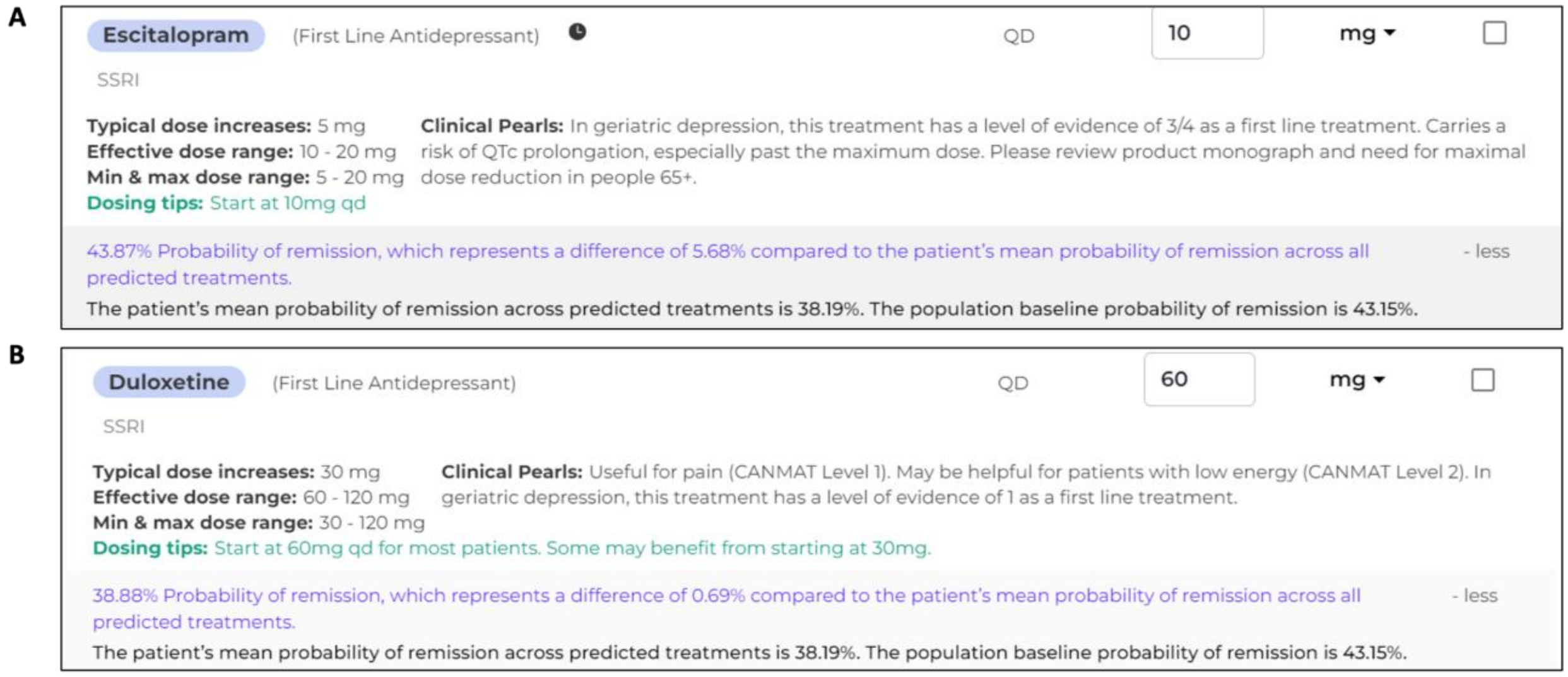

Figure 5.1 - Example result for A) escitalopram and B) prediction. Information displayed includes drug class, typical dose, effective dose range, minimum/maximum dose range, dosing tips, clinical pearls, patient-specific remission probability for escitalopram, patient-specific remission probability across treatments, and population baseline. [Note: Duloxetine is an SNRI; this was a known error which was intended to be fixed in a software patch to be done after this screenshot was taken].

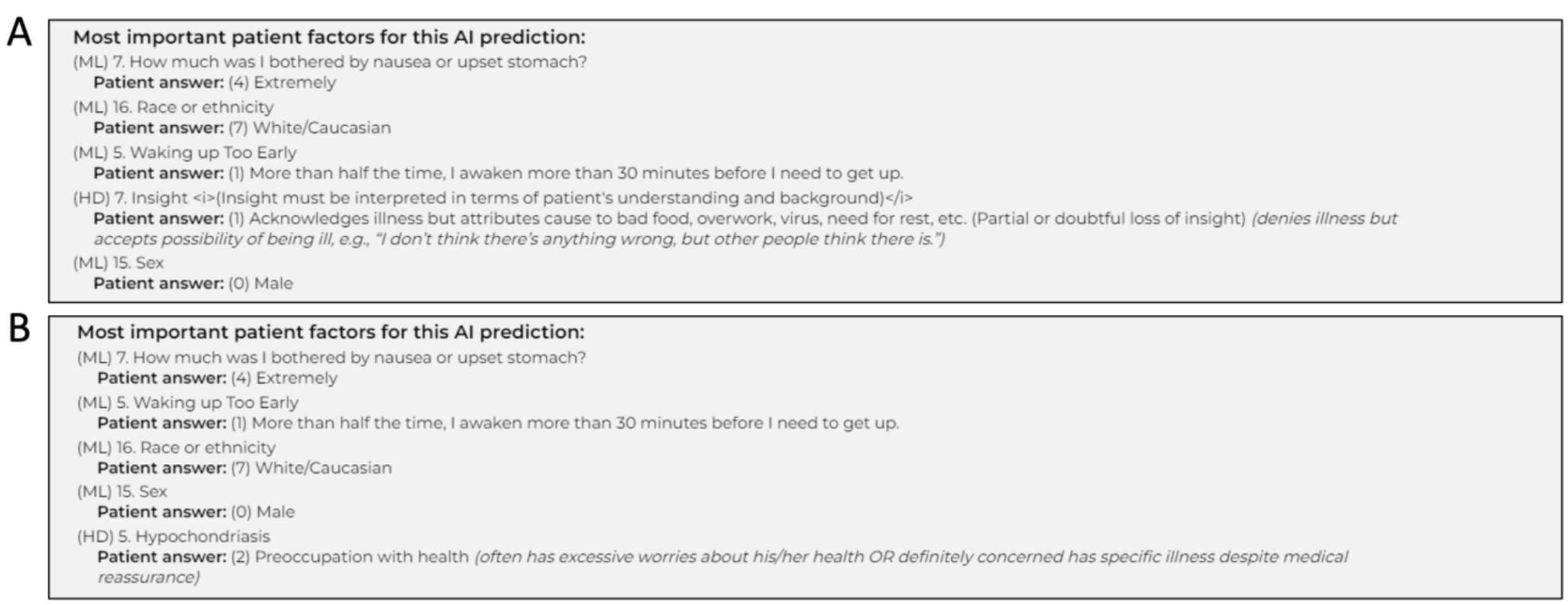

Figure 5.2 - Example interpretability report for A) Escitalopram and B) Duloxetine that provides the top 5 most salient variables for each treatment prediction. *In this example, a question related to gastrointestinal symptoms is ranked first, which is interesting given that escitalopram is known to have superior tolerability, and the patient is noting significant gastrointestinal symptoms at baseline- a situation in which an SSRI with a more favorable tolerability profile would be preferred. In the second example, for duloxetine, some features are similar (e.g.*

*gastrointestinal sensitivity) but is an interesting difference: hypochondriasis is included, which may be consistent with duloxetine's known benefit for somatic symptoms [13].*

Box 1:

This interface was developed after extensive testing and consultation with clinicians[14,20,21].

The interface provides:

- Drug name
- Dosing information
- Clinical pearls derived from the literature
- AI results (provided in purple for differentiation). As per requests from clinicians during development, these are presented in an ordered list based on the probability of remission.

Interpretability:

The platform provides:

- The mean remission rate over all 10 treatments predicted for this patient, and how they rank relative to each other.
- The baseline remission rate of 43.15% in order to help clinicians get a sense of how the individual patient compares to the sample of depressed adults used to train the model.
- The interpretability report- the 5 features which were most predictive of the outcome for that particular treatment, alongside participant responses to the question representing the feature.

CONTRIBUTIONS:

DB helped conceptualize the study, provided supervision, and wrote and revised the manuscript. RF, JM, CA helped conceptualize the analyses, performed the analyses and helped write the manuscript; SI reviewed the manuscript, KP helped conceptualize the taxonomy used in the dataset, and helped write the manuscript; GT provided data access and supervision. AK helped conceptualize the analyses and helped write the manuscript.

ACKNOWLEDGEMENTS AND DATA AVAILABILITY:

This manuscript reflects the views of the authors and may not reflect the opinions or views of the data sharing platforms or initial submitters of the trial data.

The authors do not have permission to share the data used in this project as it does not belong to them; those interested in using the data should contact the GlaxoSmithKline, Eli Lilly, NIMH, or the University of Pittsburgh.

All pharmaceutical data was kindly provided by GlaxoSmithKline and Eli Lilly data, via the Clinical Study Data Request (CSDR) platform. The dataset IDs for all the sets used in this model can be found in a separate publication[8].
All information about the model from that publication, including model weights and scaling factors, can be found at the following link: https://github.com/Aifred-

Health/pharma_research_model The rest of the data used for this project was kindly provided by the NIMH as well as the IRL-GREY investigators and the University of Pittsburgh. Data and/or research tools used in the preparation of this manuscript were obtained from the National Institute of Mental Health (NIMH) Data Archive (NDA).NDA is a unified informatics platform generated and maintained by the National Institutes of Health to provide a national, publicly available resource to facilitate and accelerate research in mental health, and support the open science movement. Dataset identifier(s): Sequenced Treatment Alternatives to Relieve Depression (STAR*D) #2148, Combining Medications to Enhance Depression Outcomes (CO-MED) #2158, Research Evaluating the Value of Augmenting Medication with Psychotherapy (REVAMP) #2153, Establishing Moderators/Biosignatures of Antidepressant Response - Clinical Care (EMBARC) MDD Treatment and Controls #2199.

Code availability: Previous versions of the pipeline used to construct the model are available here: The old Vulcan (https://github.com/Aifred-Health/VulcanAI) and a full version of the model trained used the pharmaceutical data is available here: (https://github.com/Aifred-Health/pharma_research_model). Code describing model evaluation of generation of results is available in the Supplementary Material. The full protocol is proprietary, but its major elements are reproduced in this paper. This study was not registered but the accompanying clinical trial was. People with lived experience of anxiety and depression were involved in the creation of the models described.

FUNDING AND ROLE OF THE FUNDING SOURCE:

Funding was provided by Aifred Health; the ERA PERMED Vision 2020 grant supporting IMADAPT; MEDTEQ; The National Research Council via the IRAP program; the MITACS program; Investissement Quebec; and the Quebec Ministry of Economy and Innovation. The primary researchers, with the exception of GT, were employees and/or shareholders of Aifred Health and produced this research in the context of their work for Aifred Health. No other funder had a role in the development or reporting of this research.

CONFLICT OF INTEREST:

DB, CA, KP, RF, JM, SI were employees and/or shareholders of Aifred Health and supported this research in the context of their work for Aifred Health. DB receives a salary award from the Fonds de recherche du Québec – Santé (FRQS). AK, SP have previously received an honorarium from Aifred Health. JFK has been provided options in Aifred Health.

**SUPPLEMENTARY:**

**Further details on study selection:**

Studies included were not restricted based on whether the study was successful or its endpoints were met, but we did not include participants from studies who were taking experimental drugs which were later not approved by regulatory agencies and therefore not available on the market. In general, if a multi-arm study included participants in one or more arms who matched our intended use population (IUP), we included participants from those arms but not from the arms that did not match the IUP.

With respect to sample size, sample size was deemed to be sufficient given similar results to previous, smaller studies.

**Further information on naive and conservative testing:**

Our first “naïve” experiment aims at estimating the performance of the model if it was applied to a hypothetical “new” clinical population. To create a test set we used the 875 subjects from the test dataset. To get a probability of remission, we pass each subject in the held-out sample, paired with 10 drugs separately, through the final model, once for each possible drug in the dataset. The output of the forward pass is the probability of remission for that subject for that given drug. For each of the 875 subjects we take the drug with the highest probability of remission and obtain the mean remission rate of those subjects. Finally, we take the difference between the mean remission rate of the entire dataset and our mean remission rate for the test subjects. We estimate the remission probability when using our neural network to assign patients to drugs when compared with the baseline study drug assignment. In this naïve version of the analysis, we look at hypothetical cases in which we do not necessarily know the outcome of giving a certain patient a drug. We also assess the mean difference between the best and worst predicted remission rate for each patient, to provide a sense of the spread between treatment remission probabilities as this was often noted by clinicians as being important (e.g. an accurate model that only finds a 5% difference between the best and worst drug was not seen as being clinically useful by some clinicians consulted[14]).

Our second “conservative” analysis is derived from the method established by Kapelner *et. al.* [46] and only considers the non-hypothetical cases in which we know the outcome of giving a specific treatment option to a patient[7,18]. This analysis answers the question: does our deep learning neural network personalization model outperform chance allocation? We used 100 independent and identically distributed bootstrap samples. In creating these new samples, we resample from the original dataset with replacement, train a model several times using 10-fold cross validation, and compute an “improvement score” that compares the chance remission rate with the improved remission rate found using the personalized assignations from our DNN model. To exclude hypothetical cases, we compare improvement scores between patients who actually received (by chance) the drug our model predicted they should have received to the rest of the study population. This allows us to generate a distribution, over the bootstrapped samples, of remission rates for subjects receiving model-optimized treatment versus those who do not. This, in turn, lets us compare these two populations to determine if their mean predicted remission rates are statistically significantly different from each other, and if the actual population remission rates are higher for those obtaining model-coherent treatment. Taken together, the results of the naïve and conservative analyses act to help define a range of potential effect sizes which the model could be expected to produce in clinic[7].

For conservative testing, note that on average, roughly 713.75 (SD = 80.2, max = 933, min = 554) patients were predicted as having optimal treatment in the bootstrap folds, resulting in a ~9% rate of optimal treatment by chance, in line with expectations. As the total number of patients receiving escitalopram in the bootstrap dataset should be roughly 494, this indicates that in most versions of the model, escitalopram was not the sole optimal treatment selected. As such, with more and different data, as expected in the future, more variability in the number one optimal treatment would be expected.

**Model merging:**

Due to contractual obligations, data from pharmaceutical companies needed to be kept separate from data from other sources during training. As such, models were trained and their performance was optimized using the validation datasets within each data source separately. The best performing architectures were selected and then merged to produce a final model. This final merged model was then run on the test set, which was comprised of data from both data sources. This test set was not used in any way during model training. The results of one of these models, the model trained on data provided through data requests made to pharmaceutical companies, are published in Perlman et al., 2024. A previous version of the model trained on data provided by the NIMH is available in Mehltretter et. al 2019.

**Supp. Table 1: Final n per drug:**

| | n |
|---|---|
| Citalopram | 2311 |
| Paroxetine | 1502 |
| Duloxetine | 1428 |
| Venlafaxine | 988 |
| Fluoxetine | 662 |
| Bupropion | 628 |
| Sertraline | 606 |
| Escitalopram | 544 |
| Venlafaxine-XR + Mirtazapine | 188 |
| Escitalopram+Bupropi | 185 |
| on | |

**Further Early Stopping Details:**

Within a training iteration, we make use of Early Stopping which is responsible for checkpointing model states and preserving optimal generalizability of the proposed model by tracking the validation accuracy within a training interval. If after 3 full training iterations (e.g. a patience of 3 epochs) the model is seen to have a steady drop in validation accuracy while the training accuracy continues to improve, this indicates that the model is beginning to overfit the data. Through preliminary experiments with Bayesian Optimization where the number of epochs is an attribute of variation from one experiment to another, then we can localize a range of training epochs that leads to model convergence. This range can then inform the max epochs allowed for an individual training experiment (within that class of model) and the Early Stopping procedure will fine tune the viable trainable epoch range to maximize the models exposure to the data while still preserving generalizability. Our range for an individual model to converge was set between 150-300 epochs.

**Further details on model usage during inference:**

As the model is an integral component of the encompassing application used to mediate the interaction between patient, physician, and model, the following are the steps taken to take a single patient profile and get the results.

With each request of the model, the JSON objects are first directed to the preprocessing function, where the first step is equipercentile equating, where several of the questions answered are transformed using a pre-defined dictionary into values that match our equipercentile equating process (i.e., the transformation of question response values into one standard for each “group” of similar questions).

We then one-hot encode categorical questions into the values that are expected by the model (e.g. Such as the mapping of Gender where Female (1) and Male (0)). Then the values are standardized  and add an imputation constant for those variables that went through the imputation process. The process of imputation led to some of the feature values to be zero which made the training process unstable, thus, we added in a constant uniformly to all the imputed features such to avoid a large amount of zeros in the training set which led to better model training stability. A constant was added to imputed values during training to ensure no value was at 0 to aid the learning process. This was followed by completing forward passes of the model and then presenting the results to the clinician as discussed in the main paper.

No missing data is allowed to be passed to the model using inference due to the manner in which the interface is set up.

**Further information on calibration:**

Since our data doesn't allow our models to extend their prediction certainties into the outer bounds of confidence (close to 0 or 100%), we limited our calibration evaluation towards less extreme boundaries of remission likelihood.

**Information on Prespecified metrics:**

We included 2 pre-specified metrics of model performance which were required to be met for a model to be selected as the final model after testing on the held-out test set. The first was a minimum AUC of 0.65 on the held-out test set. This number was chosen with the following reasoning. Previous models we and others have produced had AUCs close to or near 0.7 [6,7], and these models had previously been predicted to potentially improve treatment outcomes. The current dataset, while larger in terms of total number of subjects and available treatments than any model we have previously produced, also had fewer available high-quality features going into feature selection than our earlier models. This was a result of the need to use the common available features across disparate studies, and we were aware that certain metrics, such as education and income, which we had previously shown to be predictive of outcome [6] were no longer available. As such, we expected that there might be a reduction in AUC below previous results; however, results lower than 0.65 would not be likely to be clinically useful. Therefore, 0.65 was selected as a realistic estimate that would still provide benefit over chance allocation. The second pre-specified metric was that in the bias testing we would not tolerate more than a 5% underprediction of remission for any age, sex, or race/ethnicity subgroup, in order to ensure that the model did not learn to amplify harmful biases for any sociodemographic subgroup[8].

**Further notes on Bias Testing:**

In some cases, if the sample sizes in the test or validation set of a particular drug are small, sampling effects may lead to lower or higher than expected remission rates for a given drug. In these situations, to minimize the potential bias effects, we were satisfied if the model's performance approximated the mean predicted remission rate, since it has been shown that most drugs should perform roughly equally well at the population level (within margins of error) [12].

**Testing for Study Level Variables:**

As the studies varied, for example with respect to their design and their inclusion/exclusion criteria, study level variables were assessed to determine whether they were more predictive of remission than individual patient-level variables. It is important to note that for each study included there we many (sometimes over a dozen or more) sites; as such, an analysis of site effects is impossible, especially as there was no way to link sites between studies. We found that the majority of explained variance was derived from individual patient-level variables.

In order to determine that the study design elements (e.g. randomization, design, study location, etc…) do not act as confounders, we conducted hierarchical regression tests. We first created a regression model with all candidate variables for our machine learning model – those at an individual patient level, and then examined the difference between models when we additionally added for each patient the relevant study design element variables. The process used for this was to use SPSS's logistic regression function (In SPSS, Analyze>Regression>Binary Logistic; Bootstrap was set to 'on' with 1000 samples; all other options left as standard. This was performed once for the individual level patient data, and once for the individual level data plus the study level data added for each participant. This resulted in nested models which could be compared to determine if the larger model produced meaningful changes in performance). The null model produced a classification accuracy of 56.8%. The model with patient level variables only produced a classification accuracy of 62.7% and a Naegelkerke pseudo r2 of 0.1. The model with patient level variables and the study level variables produced a classification accuracy of 64.6% and a Naegelkerke pseudo r2 of 0.14. As such we found a very small percentage improvement in the accuracy and pseudo r2 of the second model, much smaller than the effect of the individual patient variables model compared to a null model (predict the majority class). As such, study level variables have a small but likely not clinically significant impact on our results. We can further review the study level variables which were significant to determine if there were confounds which make their true nature as study level variables rather than drug-or-patient specific variables less certain, further weakening the influence of study level variables. A study being conducted in Western Europe had a beta –6.7 and a p value of 0.012 (note this would not be significant at an alpha of 0.01 which may be more conservative given the large number of features used). Upon further review, the 541 patients in the dataset from Western Europe were from two studies, one of which had 51% of its patients assigned to bupropion, and the other had 42% of their subjects assigned to bupropion; given that bupropion was one of the least performing drugs in terms of remission rate, this may provide a partial explanation of the lower remission rate for Western Europe. Blinding had a positive beta (4.4) and a p value of 0.003; this is likely explained by the fact that the largest unblinded study was STAR*D, which had one of the antidepressants of middle efficacy as per metanalyses (citalopram) and more patients with chronic or recurrent depression than other studies. As such, this variable likely was picking up on patient population and drug differences, which are accounted for patient variable model, rather than being truly about study design. Whether a study had a placebo lead-out did have a beta of 1.1 and p value of 0.005; the studies with placebo lead out did have an overrepresentation of one of the higher performing medications (duloxetine 62%) compared to studies without placebo lead out (where adding together venlafaxine, duloxetine, escitalopram, venlafaxine-mirtazapine, escitalopram-bupropion and sertraline, the highest performing drugs, we get 41%). This over-representation of a more effective drug may be a confound which at least partially explains the placebo lead-out advantage. Finally, years since study start has a beta of –0.6, with a p value of 0.03. This result is likely explained by the fact that older antidepressants in the SSRI/SNRI/NDRI categories (such as paroxetine and fluoxetine) are less effective than newer ones (such as escitalopram for example) as per guidelines and some metanalyses and this was certainly true in our data; as such this variable is likely capturing drug effectiveness at the group level, and drug is accounted for in the individual patient data model. As such, only placebo lead out and blinding would survive quasi-robust correction of p-values and also represents a study level variable, and in all cases the details about the drugs used in the studies seems to be a confounding variable which may explain part of the effect. While whether or not treatment resistant patients were allowed in the study was not a variable selected by the regression model (likely as it co-varied with other study characteristics), including it in model by force did not appreciably change the results.

**Supplementary Figure 1: Train Set AUC Plots**

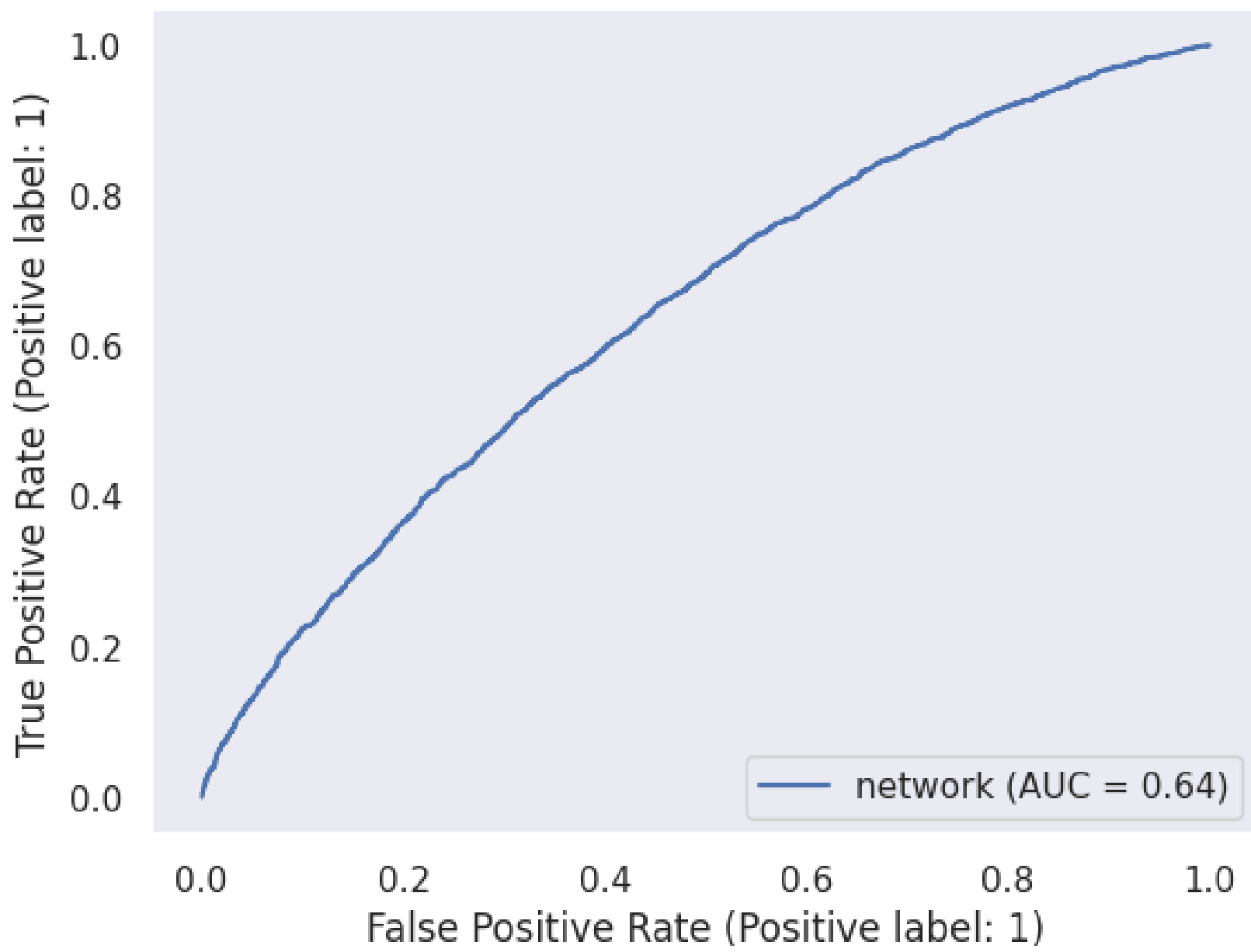

**Supp. Fig. 2 Test Set AUC Plot**

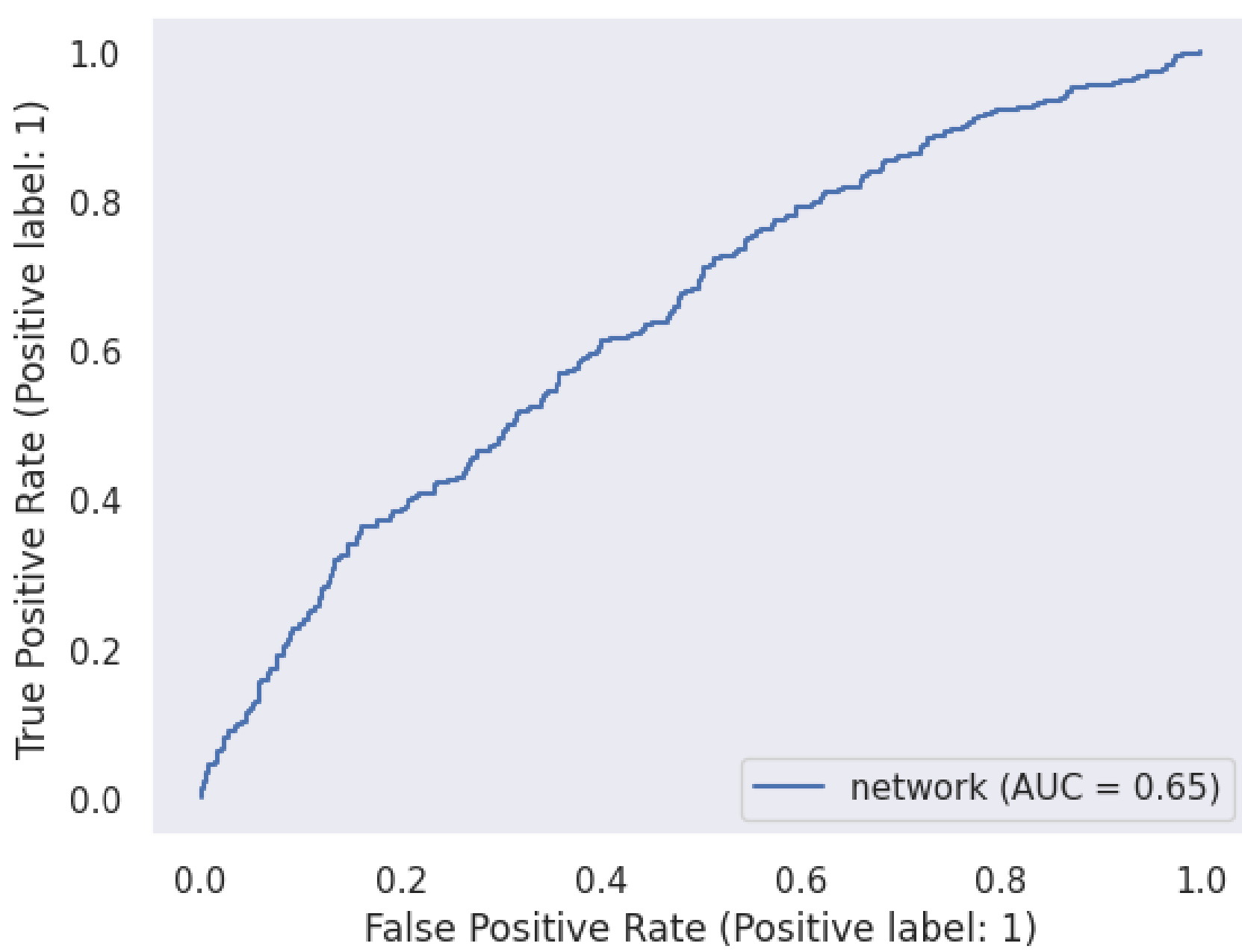

**Supp. Fig 3. Validation Set AUC Plots**

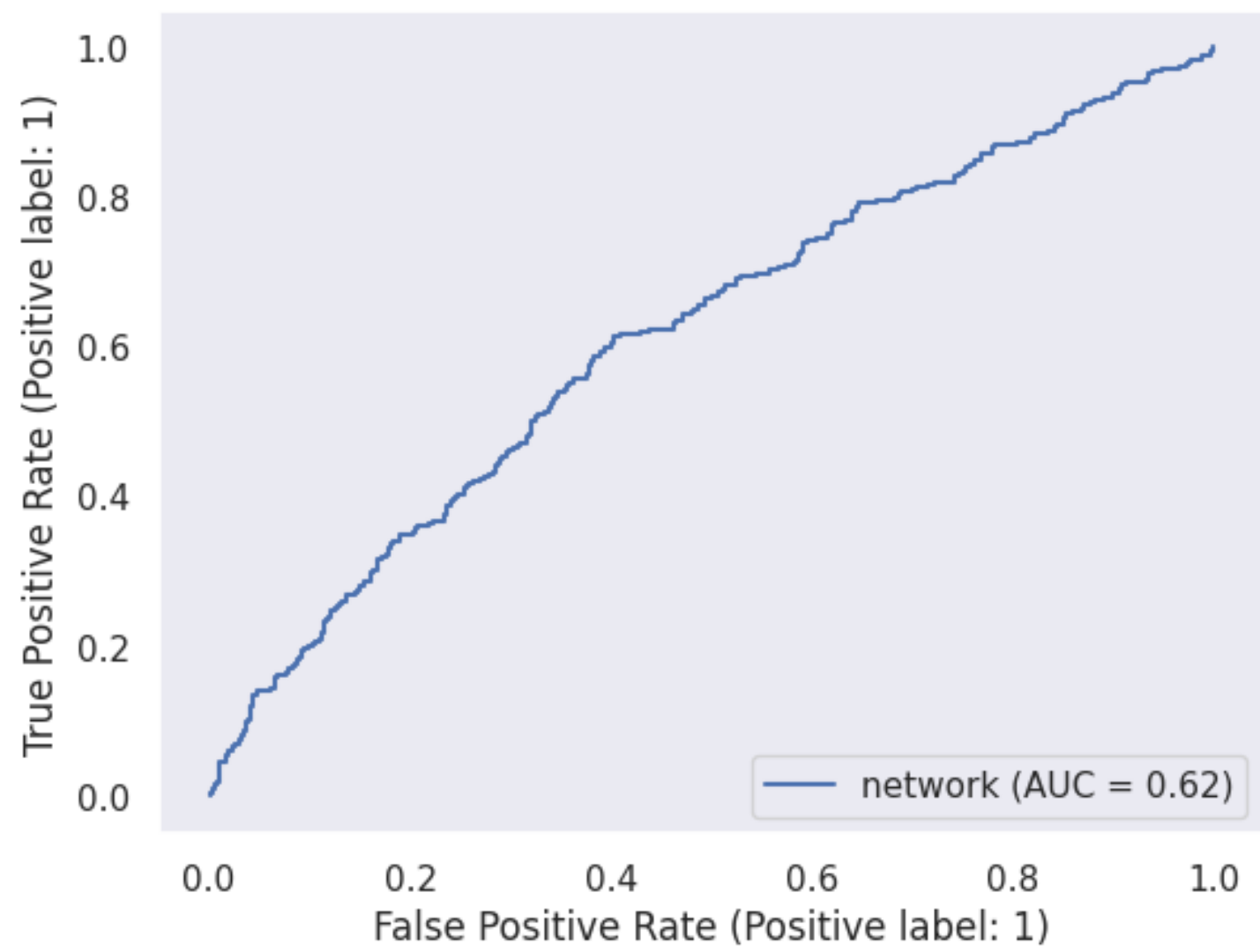

**Model Accuracy Results by Treatment Subgroup:**

Note: limited sample size was available for several treatments; future updates of this model with more data for treatments with limited n is expected to improve treatment subgroup performance. We note that in the training set AUC's of 0.6 or more were achieved for all drugs except sertraline, where the AUC was close to 0.6. Results in the test and validation set were more variable; we note that there were low sample sizes for some of the tested drugs in these sets.

**Supp. Table 2: Model Accuracy Results by Treatment Subgroup, Train Set**

| | Accuracy | Sensitivity | Specificity | PPV | NPV | F1 | AUC |
|---|---|---|---|---|---|---|---|
| Bupropion (516) | 0.6335 | 0.132 | 0.932 | 0.5 | 0.68 | 0.21 | 0.644 |
| Escitalopram + Bupropion (147) | 0.61 | 0.485 | 0.727 | 0.62 | 0.61 | 0.54 | 0.66 |
| Duloxetine (1160) | 0.558 | 0.789 | 0.308 | 0.552 | 0.575 | 0.65 | 0.603 |
| Escitalopram (446) | 0.581 | 0.827 | 0.269 | 0.588 | 0.552 | 0.687 | 0.627 |
| Fluoxetine (535) | 0.671 | 0.355 | 0.813 | 0.461 | 0.737 | 0.401 | 0.647 |
| Venlafaxine-XR + Mirtazapine (152) | 0.585 | 0.492 | 0.655 | 0.516 | 0.633 | 0.503 | 0.602 |
| Citalopram (1873) | 0.594 | 0.508 | 0.664 | 0.551 | 0.624 | 0.529 | 0.627 |
| Paroxetine (1217) | 0.653 | 0.444 | 0.779 | 0.549 | 0.698 | 0.491 | 0.66 |
| Venlafaxine (812) | 0.551 | 0.7 | 0.424 | 0.509 | 0.624 | 0.59 | 0.611 |
| Sertraline (492) | 0.581 | 0.44 | 0.657 | 0.413 | 0.683 | 0.427 | 0.586 |

**Supp. Table 3: Model Accuracy Results by Treatment Subgroup, Test Set**

| | Accuracy | Sensitivity | Specificity | PPV | NPV | F1 | AUC |
|---|---|---|---|---|---|---|---|
| Bupropion (60) | 0.633 | 0.157 | 0.853 | 0.333 | 0.686 | 0.214 | 0.49 |
| Escitalopram + Bupropion (18) | 0.5 | 0.222 | 0.777 | 0.5 | 0.5 | 0.307 | 0.703 |
| Duloxetine (149) | 0.69 | 0.847 | 0.48 | 0.685 | 0.704 | 0.757 | 0.703 |
| Escitalopram (50) | 0.58 | 0.851 | 0.26 | 0.575 | 0.6 | 0.686 | 0.63 |
| Fluoxetine (67) | 0.686 | 0.381 | 0.826 | 0.5 | 0.745 | 0.432 | 0.687 |
| Venlafaxine-XR + Mirtazapine (18) | 0.44 | 0 | 0.727 | 0 | 0.53 | 0 | 0.53 |
| Citalopram (212) | 0.561 | 0.26 | 0.81 | 0.532 | 0.569 | 0.349 | 0.591 |
| Paroxetine (154) | 0.655 | 0.41 | 0.795 | 0.534 | 0.702 | 0.464 | 0.65 |
| Venlafaxine (93) | 0.505 | 0.536 | 0.481 | 0.448 | 0.568 | 0.488 | 0.541 |
| Sertraline (54) | 0.537 | 0.211 | 0.714 | 0.286 | 0.625 | 0.24 | 0.58 |

**Supp. Table 4: Model Accuracy Results by Treatment Subgroup, Validation Set**

| | Accuracy | Sensitivity | Specificity | PPV | NPV | F1 | AUC |
|---|---|---|---|---|---|---|---|
| Bupropion (52) | 0.615 | 0.181 | 0.933 | 0.666 | 0.608 | 0.285 | 0.454 |
| Escitalopram + Bupropion (20) | 0.55 | 0.333 | 0.727 | 0.5 | 0.57 | 0.4 | 0.565 |
| Duloxetine (119) | 0.672 | 0.841 | 0.482 | 0.646 | 0.729 | 0.731 | 0.687 |
| Escitalopram (48) | 0.562 | 0.884 | 0.181 | 0.561 | 0.571 | 0.686 | 0.634 |
| Fluoxetine (60) | 0.65 | 0.263 | 0.829 | 0.416 | 0.708 | 0.322 | 0.612 |
| Venlafaxine-XR + Mirtazapine (18) | 0.55 | 0.625 | 0.5 | 0.5 | 0.625 | 0.555 | 0.487 |
| Citalopram (226) | 0.601 | 0.561 | 0.632 | 0.539 | 0.653 | 0.549 | 0.641 |
| Paroxetine (131) | 0.603 | 0.34 | 0.75 | 0.432 | 0.67 | 0.38 | 0.568 |
| Venlafaxine (83) | 0.506 | 0.722 | 0.34 | 0.456 | 0.615 | 0.559 | 0.53 |
| Sertraline (60) | 0.55 | 0.333 | 0.666 | 0.35 | 0.649 | 0.341 | 0.515 |

**Supp. Figure 4: Train Set Calibration Plots**

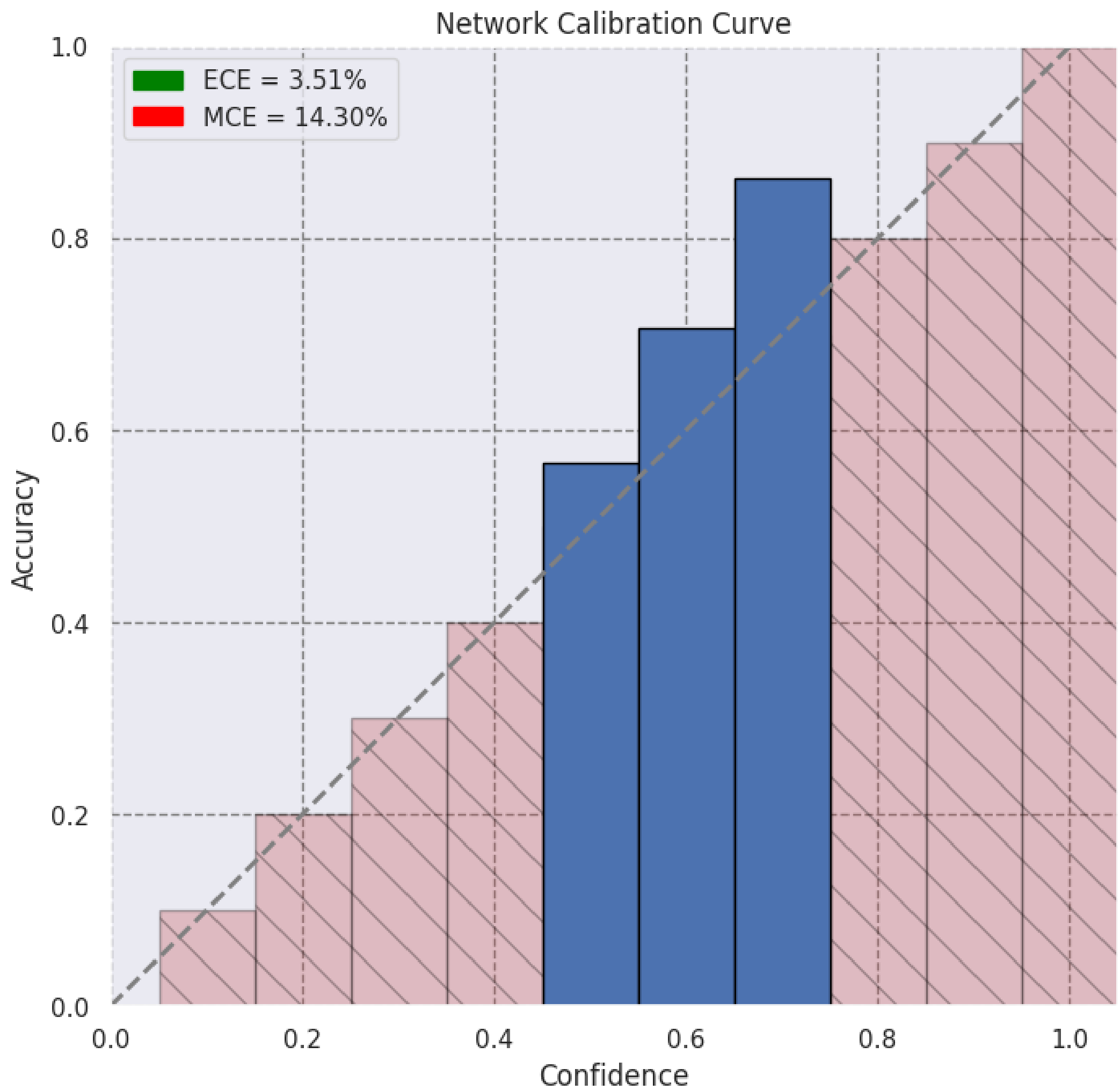

**Supp Figure 5: Validation Set Calibration Plots**

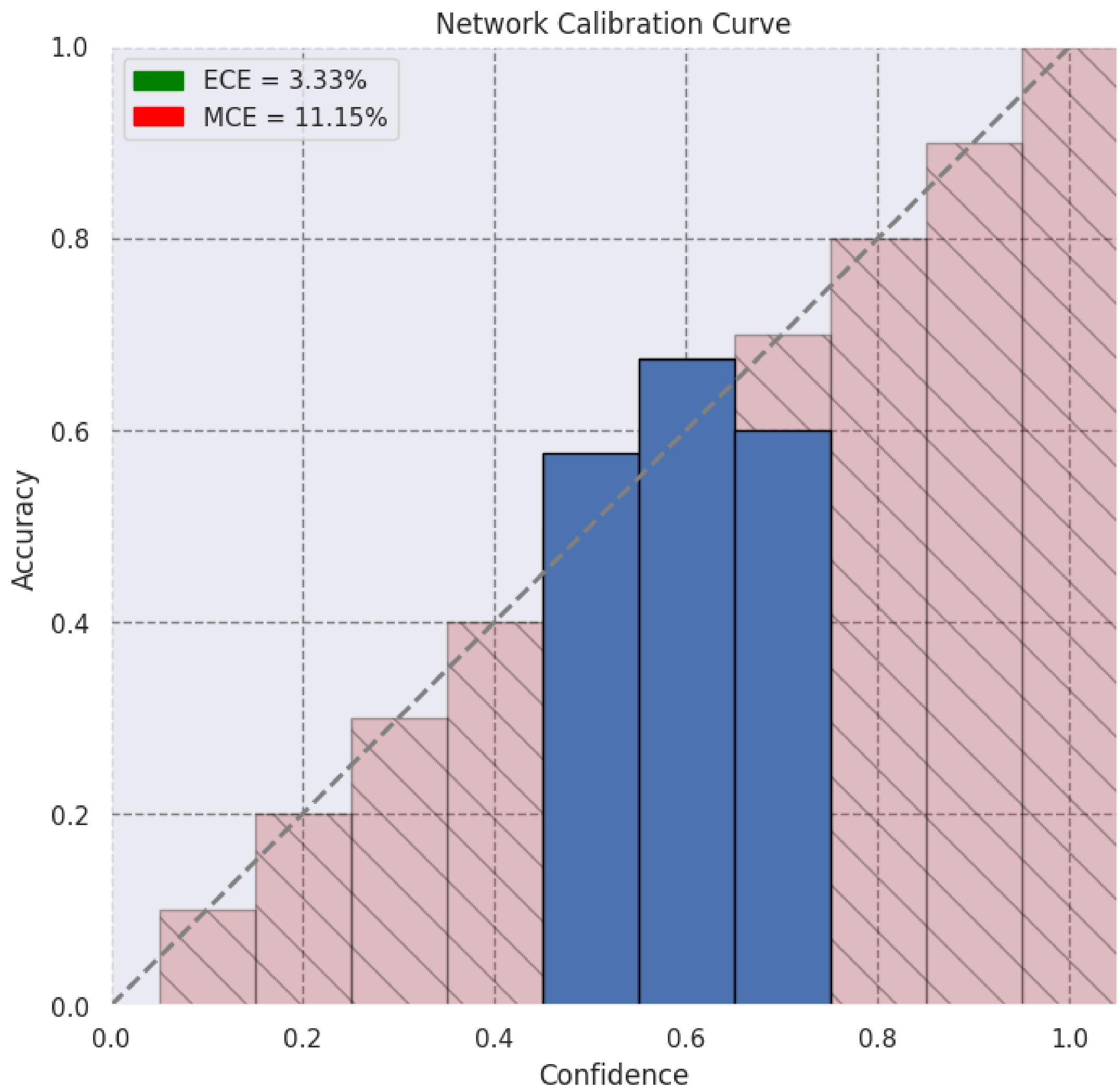

**Supp. Figure 6: Train Set Sensitivity Plots**

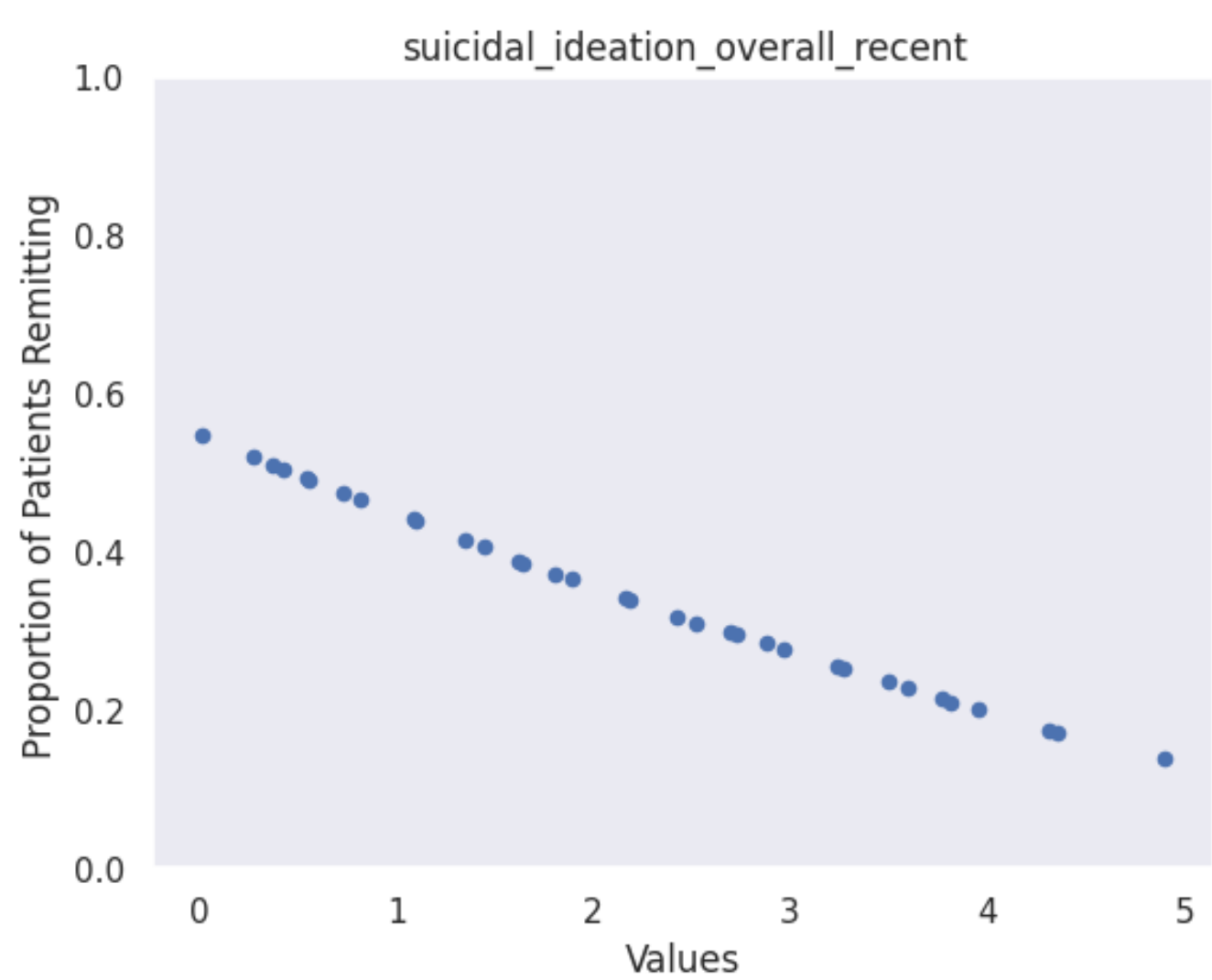

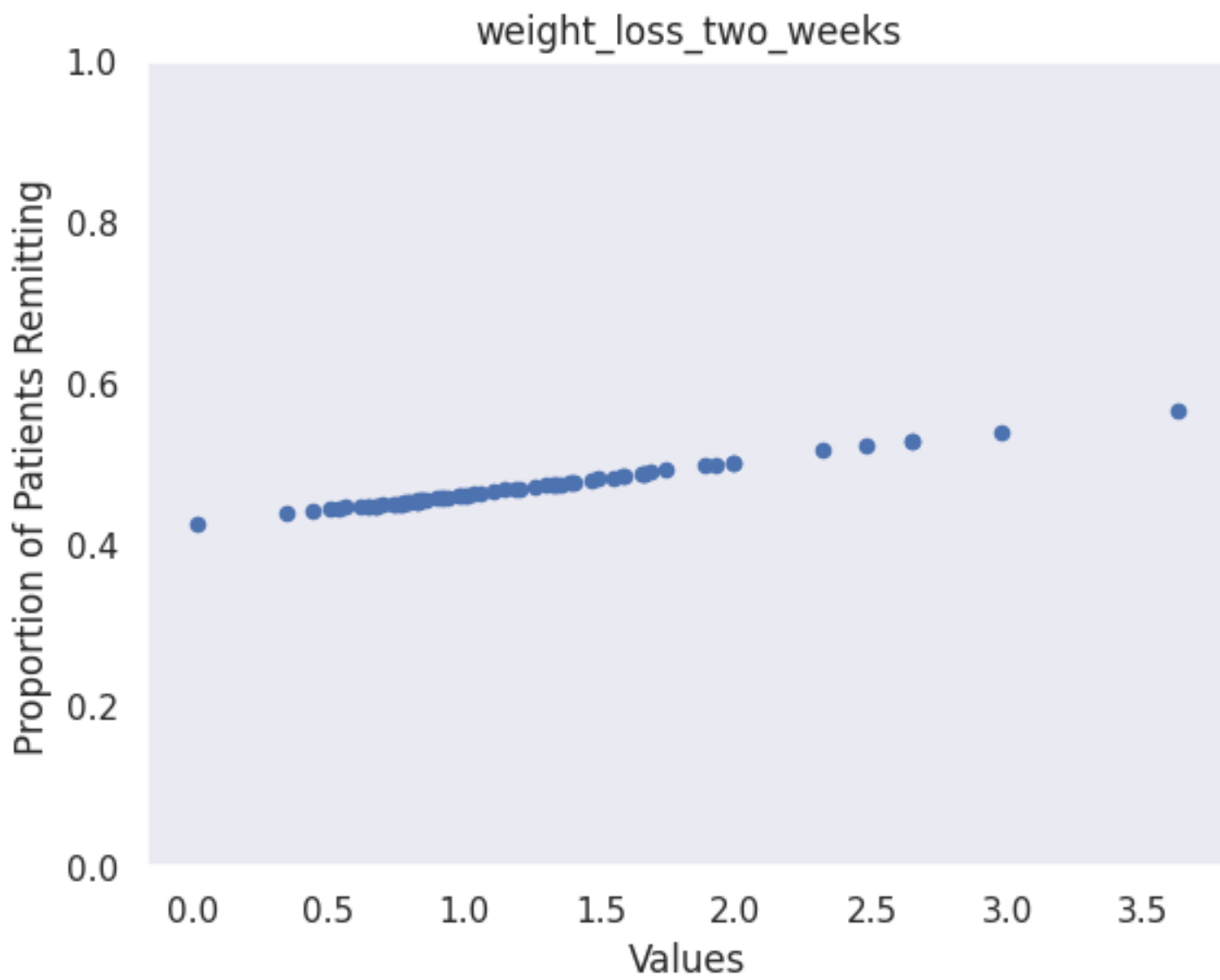

weight_loss_two_weeks
1.0
0.8
0.6
0.4
0.2
0.0
Proportion of Patients Remitting
0.0
0.5
1.0
1.5
2.0
2.5
3.0
3.5
Values

**Supp. Figure 7: Validation Set Sensitivity Plot**

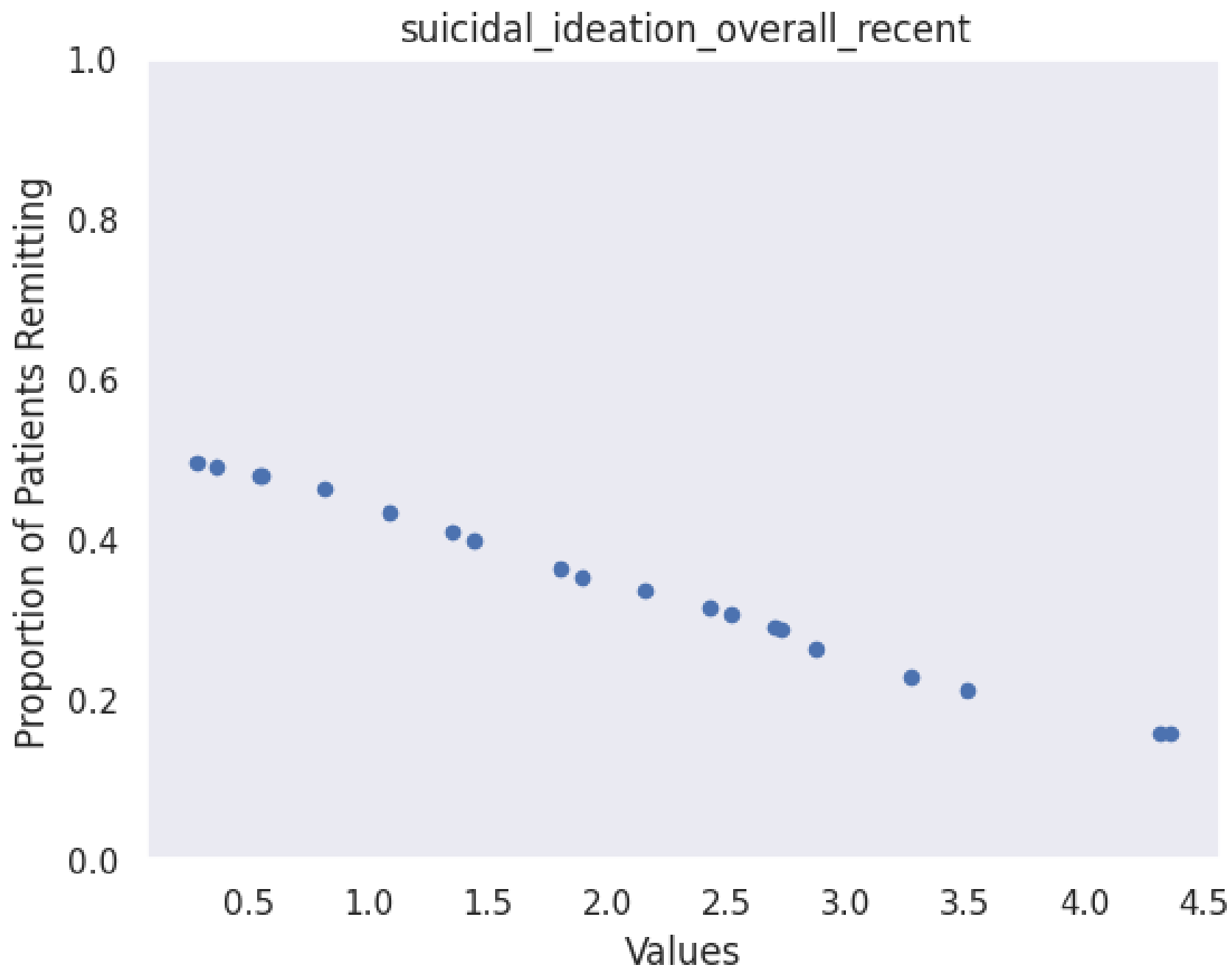

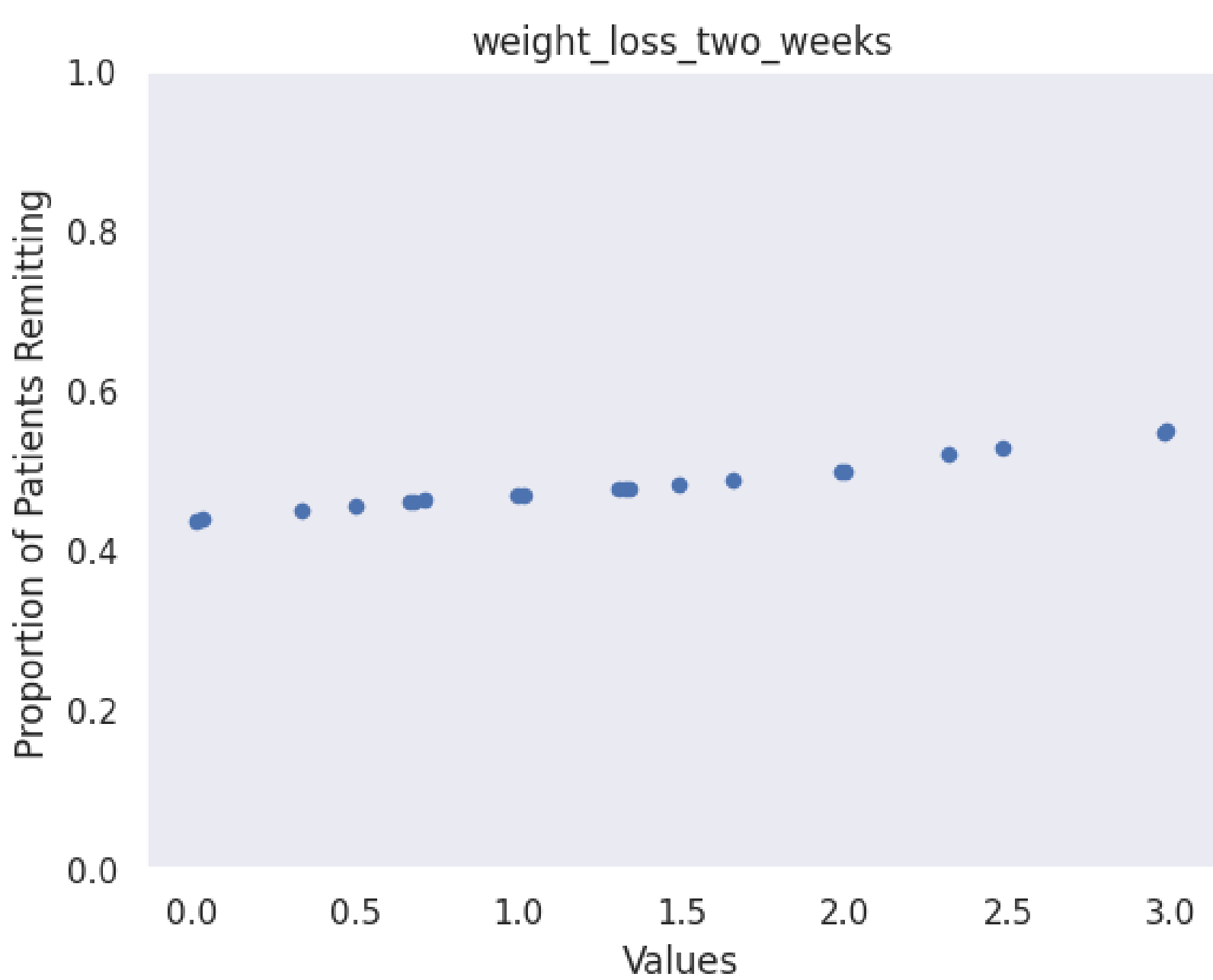

**Differential treatment rankings**

**Supp. Table 5: Differential Treatment Ranking on Test Set**

| | Rank 1 | Rank 2 | Rank 3 | Rank 4 | Rank 5 | Rank 6 | Rank 7 | Rank 8 | Rank 9 | Rank 10 |
|---|---|---|---|---|---|---|---|---|---|---|
| Bupropion | 0 | 0 | 0 | 0 | 0 | 2 | 5 | 61 | 142 | 665 |
| Citalopram | 0 | 46 | 144 | 138 | 185 | 305 | 56 | 1 | 0 | 0 |
| Escitalopram | 860 | 13 | 2 | 0 | 0 | 0 | 0 | 0 | 0 | 0 |
| Escitalopram + Bupropion | 2 | 15 | 63 | 157 | 230 | 206 | 181 | 17 | 4 | 0 |
| Sertraline | 0 | 7 | 27 | 64 | 86 | 129 | 460 | 79 | 17 | 6 |
| Venlafaxine | 5 | 368 | 206 | 135 | 108 | 37 | 15 | 1 | 0 | 0 |
| Venlafaxine-XR + Mirtazapine | 7 | 184 | 194 | 168 | 138 | 129 | 46 | 9 | 0 | 0 |
| Duloxetine | 1 | 242 | 239 | 213 | 122 | 55 | 2 | 1 | 0 | 0 |
| Fluoxetine | 0 | 0 | 0 | 0 | 4 | 10 | 69 | 485 | 196 | 111 |
| Paroxetine | 0 | 0 | 0 | 0 | 2 | 2 | 41 | 221 | 516 | 93 |

**Supp Table 6: Differential Treatment Ranking on Validation Set**

| | Rank 1 | Rank 2 | Rank 3 | Rank 4 | Rank 5 | Rank 6 | Rank 7 | Rank 8 | Rank 9 | Rank 10 |
|---|---|---|---|---|---|---|---|---|---|---|
| Bupropion | 0 | 0 | 0 | 0 | 1 | 1 | 5 | 59 | 134 | 617 |
| Citalopram | 0 | 67 | 115 | 149 | 146 | 293 | 44 | 3 | 0 | 0 |
| Escitalopram | 790 | 20 | 6 | 0 | 1 | 0 | 0 | 0 | 0 | 0 |
| Escitalopram + Bupropion | 0 | 16 | 76 | 142 | 251 | 180 | 144 | 7 | 1 | 0 |
| Sertraline | 0 | 5 | 11 | 46 | 78 | 112 | 431 | 102 | 27 | 5 |
| Venlafaxine | 1 | 319 | 219 | 153 | 72 | 41 | 12 | 0 | 0 | 0 |
| Venlafaxine-XR + Mirtazapine | 25 | 189 | 210 | 123 | 112 | 118 | 32 | 8 | 0 | 0 |
| Duloxetine | 1 | 201 | 180 | 203 | 156 | 68 | 8 | 0 | 0 | 0 |
| Fluoxetine | 0 | 0 | 0 | 0 | 0 | 3 | 92 | 419 | 189 | 114 |
| Paroxetine | 0 | 0 | 0 | 1 | 0 | 1 | 49 | 219 | 466 | 81 |

### *Comparison to null model*

We explored accuracy p-values compared to a null model (i.e., a model predicting the majority class which, for all our datasets, resulted in a comparison to a null model accuracy that would only predict no remission, compared using a one-sided binomial test). We found that our model outperformed a null model on the training set ($p = 1.2e-09$); did not outperform the null, but was approaching significance, in the validation set ($p = 0.08$); and outperformed the null in the test set ($p = 0.01$).

**Software package details:**

The following is a list of the open-source Python packages used throughout the model building, testing, and deployment phases of the project. These packages are integral to various aspects of the machine learning pipeline, including data manipulation, model training, evaluation, and visualization.:

**Core Data Manipulation and Computation**

- **numpy>=1.21.1:** Essential for numerical operations and array manipulation.
- **pandas>=1.5.0:** Used for data manipulation and analysis, particularly with structured data.
- **scipy>=1.7.1:** Extends numpy with additional mathematical and scientific functionality.
- **pydash>=5.0.2:** Utility library for working with collections, strings, and numbers.

**Machine Learning and Model Building**

- **scikit-learn>=1.0:** Provides tools for classical machine learning algorithms (e.g., regression, classification, clustering).
- **torch>=1.9.0:** Core library for deep learning, providing tools to build and train neural networks.
- **torchvision>=0.10.0:** Extends PyTorch with vision-specific functionalities (image transformations, pre-trained models).
- **ax-platform>=0.2.8:** Optimization framework for hyperparameter tuning, typically used for model optimization.

**Model Interpretation and Testing**

- **skater==1.0.4**: Model interpretation tool, providing explainability for complex models like deep learning.
- **pytest>=6.2.4**: Testing framework to ensure correctness and quality of code and models.

**Visualization**

- **matplotlib>=3.4.2**: Core library for creating static visualizations and plots.
- **seaborn>=0.12**: Built on top of matplotlib, it simplifies statistical visualizations with enhanced aesthetics.

**Package Management and Documentation**

- **setuptools>=57.4.0**: Manages packaging and distribution of Python code.
- **sphinx>=4.1.2**: Documentation generator for Python projects, making it easier to create readable project docs.

**Data Preprocessing and Imputation**

- **optbinning==0.12.1**: For optimal binning of data, useful for feature engineering and preprocessing.
- **autoimpute==0.13.0**: Handles missing data imputation with advanced techniques like matrix completion.

**Utilities and Development Tools**

- **ipython>=7.26.0**: Enhanced interactive Python shell for easier debugging and exploration.
- **tqdm>=4.62.0**: Progress bar library for tracking task progress, useful during long-running processes.

These technologies and packages combine to form a robust infrastructure that ensures the ML model is built, trained, tested, and deployed in a scalable, efficient, and reproducible manner. Each tool and library plays a role in the overall machine learning pipeline, from model design to deployment in production.

Final Model Accuracy Results. Columns 4-8 results are based on classifying remission if the predicted remission probability is 50% or greater.

| Data Source | Accuracy | AUC | F1-Score | Sensitivity | Specificity | NPV | PPV |
|---|---|---|---|---|---|---|---|
| Train Set | 0.60 | 0.64 | 0.55 | 0.57 | 0.63 | 0.66 | 0.54 |
| Validation Set | 0.60 | 0.62 | 0.55 | 0.56 | 0.62 | 0.65 | 0.53 |
| Held-Out Test Set | 0.60 | 0.65 | 0.51 | 0.47 | 0.70 | 0.55 | 0.64 |